\documentclass[10pt,american,english]{article}
\usepackage[T1]{fontenc}
\usepackage[utf8]{luainputenc}
\usepackage{geometry}
\usepackage{float}
\usepackage{calc}
\usepackage{amsmath}
\usepackage{amssymb}
\usepackage{graphicx}
\usepackage{setspace}
\usepackage{esint}
\makeatletter

\let\SF@@footnote\footnote
\def\footnote{\ifx\protect\@typeset@protect
    \expandafter\SF@@footnote
  \else
    \expandafter\SF@gobble@opt
  \fi
}
\expandafter\def\csname SF@gobble@opt \endcsname{\@ifnextchar[
  \SF@gobble@twobracket
  \@gobble
}
\edef\SF@gobble@opt{\noexpand\protect
  \expandafter\noexpand\csname SF@gobble@opt \endcsname}
\def\SF@gobble@twobracket[#1]#2{}

\makeatother

\usepackage{babel}
\begin{document}
\begin{onehalfspace}

\title{Relative localization of point particle interactions}
\end{onehalfspace}

\begin{onehalfspace}

\author{Jos\'{e} Ricardo Cam\~{o}es de Oliveira%
\thanks{josercamoes@gmail.com / jdeoliveira@perimeterinstitute.ca%
}\\
\textit{\small Perimeter Scholars International}\textit{\normalsize }%
\thanks{Essay submitted as partial fulfillment of the requirements for the
Perimeter Scholars International MSc degree at the University of Waterloo/Perimeter
Institute for Theoretical Physics. First draft submitted for assessment
on June 1st, 2011. Minor corrections in the current version.%
}\textit{\normalsize }\\
\textit{\normalsize }\\
\textit{\normalsize }\\
\textit{\footnotesize supervised by}\\
\\
Laurent Freidel\\
\textit{\small Perimeter Institute for Theoretical Physics}\\
\textit{\small 31 Caroline Street North}\\
\textit{\small Waterloo N2L 2Y5, Ontario, Canada}\textit{\normalsize }\\
\textit{\normalsize }\\
\textit{\normalsize }\\
}
\end{onehalfspace}
\maketitle
\begin{abstract}
\begin{onehalfspace}
We review the main concepts of the recently introduced principle of
relative locality and investigate some aspects of classical interactions
between point particles from this new perspective. We start with a
physical motivation and basic mathematical description of relative
locality and review the treatment of a system of classical point particles
in this framework. We then examine one of the unsolved problems of
this picture, the apparent ambiguities in the definition of momentum
constraints caused by a non-commutative and/or non-associative momentum
addition rule. The gamma ray burst experiment is used as an illustration.
Finally, we use the formalism of relative locality to reinterpret
the well-known multiple point particle system coupled to 2+1 Einstein
gravity, analyzing the geometry of its phase space and once again
referring to the gamma ray burst problem as an example.

\pagebreak{}
\end{onehalfspace}

\begin{doublespace}
\tableofcontents{}
\end{doublespace}

\begin{onehalfspace}
\pagebreak{}\end{onehalfspace}

\end{abstract}
\begin{onehalfspace}

\section{Introduction}
\end{onehalfspace}

\begin{onehalfspace}
The notion of relative locality was originally proposed as a response
to criticisms to Double Special Relativity (DSR) \cite{key-3,key-4},
a modification of Special Relativity proposed as a step towards quantum
gravity. DSR proposes the replacement of the Poincaré symmetry group
of SR by the $\kappa-$Poincaré group, while preserving relativity
of inertial frames. The practical consequence of the modification
is the introduction of an universal invariant energy scale (Planck
scale) along with the invariant velocity, $c$.\\
\\
There are compelling arguments to believe that 2+1 gravity is a DSR
theory. Indeed, when coupled to point particles, the theory has \cite{key-5}
\end{onehalfspace}
\begin{itemize}
\begin{onehalfspace}
\item a universal mass limit independent of the number of particles;
\item a noncommutative spacetime Poisson algebra;
\item a non-linear 3-momentum addition rule;
\item a curved 3-momentum space;\end{onehalfspace}

\end{itemize}
\begin{onehalfspace}
all of which are characteristics of a DSR theory.\\
\\
One of the main criticisms of DSR is its apparent violation of locality.
The construction used to support this claim is the study of a particle
collision: if two worldlines intersect in some inertial frame, then
there exists a class of inertial observers for which the worldlines
do \textit{not} cross \cite{key-12}. That is, the notion of locality
is no longer invariant - an interaction that is local for an observer
can be non-local for a different one - a feature that naturally raised
concerns about the theory, since it seemed to indicate a fundamental
subjectivity of reality, violating the principle of equivalence. \\
\\
The concept of relative locality was suggested to argue this is not
the case, by proposing that the violation of locality present in DSR
theories is not real, but a consequence of misinterpreting the geometry
of reality, due to taking spacetime to be an absolute, invariant entity
- an assumption motivated by our knowledge and intuition, but effectively
unwarranted. Compare this situation with the problem of simultaneity
in SR - the \foreignlanguage{american}{counterintuitive} statement
of two events occurring at the same time for one observer being located
at different times for a different one stems from abandoning the Newtonian
notion of absolute time. In the framework of relative locality, it
is the absoluteness of spacetime that is abandoned.\\
\\
Relative locality is a change of paradigm with the intention of better
understanding quantum gravity. In this context, the problem of studying
the properties of Feynman diagrams of a field theory according to
the new framework becomes of major importance. A natural first step
to approach it is to describe how our understanding of particle interactions
changes in the simpler context of classical mechanics, which is the
main topic of the present essay.
\end{onehalfspace}

\begin{onehalfspace}

\section{The principle of relative locality}
\end{onehalfspace}

\begin{onehalfspace}

\subsection{Physical concept}
\end{onehalfspace}

\begin{onehalfspace}
As observers immersed in the universe, notions of space and time are
intuitive to us. We immediately interpret the localization of objects
around us in terms of distances, and perceive the ordering of events
and the concept of cause and effect as result of the passage of time.
Perhaps because these concepts are so natural and intuitive, all physical
theories to date take them as fundamental rather than {}``emergent''
from deeper structures. But considering more carefully the mechanisms
through which we probe spacetime, we see that within them are the
seeds for a drastic change of perspective. \\
\\
Consider a simple example - measuring the length of a bar. An observer
{}``sees'' the bar by receiving photons from it - optical mechanisms
convert them in information interpretable by the brain as an image.
And, indeed, knowing the time each of two photons, one from each end
of the bar, takes to reach the eye, together with the direction in
which they were sent, one can derive the length of the bar from basic
geometric considerations.\\
\begin{figure}[H]
\begin{centering}
\includegraphics{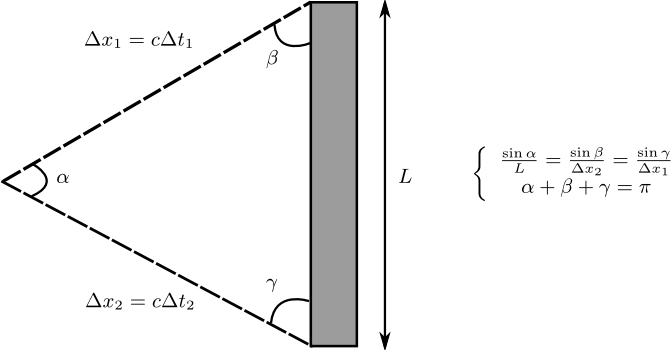}
\par\end{centering}

\caption{Deriving the length $L$ of the bar using the data $\Delta t_{1},\,\Delta t_{2},\,\alpha$
obtained from photons.}
\end{figure}
In more abstract terms, the observer performed a distance measurement
using only a calorimeter for photon detection and a clock for time
measurement. The same line of thought can be applied to other measurements
(velocity, acceleration...), in a way that suggests that the fundamental
experimental apparatuses are the calorimeter and the clock - and we
effectively probe spacetime through exchange of energy-momentum \textit{quanta}.\\
\\
Put this way, it seems almost natural to propose that the {}``arena''
where physical processes happen is not spacetime, but energy-momentum
space. Spacetime becomes a construction, an auxiliary entity derived
by observers from physical interactions. As such, there is no reason
to assume this constructed spacetime is independent of energy and
momentum (the parameters of a DSR change of frame), so in general
it can be observer-dependent - which, as will be illustrated below,
leads to locality of an interaction being itself relative to the observer.\\
\\
The change of paradigm described above leads to the formulation of
the Principle of Relative Locality. Quoting \cite{key-1}:\\
\\
\framebox{\begin{minipage}[t]{1\columnwidth}%
\textit{Physics takes place in phase space and there is no invariant
global projection that gives a description of processes in spacetime.
From their measurements local observers can construct descriptions
of particles moving and interacting in a spacetime, but different
observers construct different spacetimes, which are observer-dependent
slices of phase space. }%
\end{minipage}}
\end{onehalfspace}

\begin{onehalfspace}

\subsection{The geometry of phase space}
\end{onehalfspace}

Any theory of quantum gravity must have as fundamental quantities
Planck's constant, $\hbar$, and Newton's constant, $G$. Taking different
limits of these, we obtain different experimental regimes of gravity:
taking $\hbar\rightarrow0$ leads us to classical General Relativity,
while considering $G\rightarrow0$ should recover special-relativistic
quantum mechanics. But there is a third limit that can be taken, if
we notice that the two fundamental constants can be combined to form
the Planck mass $m_{p}=\sqrt{\frac{\mathbf{\hbar}}{G}}$: taking $G\rightarrow0\text{ and }\hbar\rightarrow0$
while keeping $m_{p}$ finite. In this regime, since the effects of
classical gravity and special-relativistic quantum mechanics are negligible,
the geometry of spacetime is Minkowski and one can work in the classical
mechanics formalism; short-distance quantum phenomena should be irrelevant
as well since the Planck length is $l_{p}=\sqrt{\hbar G}\rightarrow0.$
However, the presence of the Planck scale indicates that novel quantum
gravity effects could occur at energies of order $m_{p}$, possibly
resulting in a nontrivial geometry of momentum space.\\
\\
In line with taking momentum space $\mathcal{P}$ as fundamental,
our phase space will be its cotangent bundle, $\mathcal{T}^{*}\mathcal{P}$.
Spacetime at a given point in $\mathcal{P}$ is the cotangent space
to $\mathcal{P}$ at that point, which we denote by $\mathcal{X}(p)=\mathcal{T}_{p}^{*}\mathcal{P}$.
While we will not discuss it here, one can speculate that when $G_{N}$
and $\hbar$ are present (the hypothetical full quantum gravity theory),
the geometry of the whole phase space is something much more complex,
intertwining spacetime and momentum space in a nontrivial fiber bundle.\\

\subsubsection*{Metric of momentum space}

Our observer can construct the metric of momentum space through two
classes of measurements:
\begin{itemize}
\item Determining the rest mass of a particle gives the geodesic distance
of its position in $\mathcal{P}$ to the origin,
\begin{equation}
D^{2}(p,0)=m^{2};\label{eq:relocshell}
\end{equation}

\item On the other hand, a measurement of kinetic energy $K$ gives the
distance between the moving particle's momentum, $p$ and that of
the same particle at rest, $p'$:
\begin{equation}
D^{2}(p,p')=-2mK.
\end{equation}

\end{itemize}
This information is sufficient to recover the metric of momentum space%
\footnote{Note that, to keep in line with the convention of lower indices for
momenta, lower indices are contravariant and upper indices are covariant.%
},
\begin{equation}
dk^{2}=D^{2}(k,\, k+dk)\equiv g^{ab}(k)\, dk_{a}dk_{b}.
\end{equation}

\subsubsection*{Connection of momentum space}

Considering interaction processes between point particles, it becomes
evident that a combination rule for momenta is necessary, since it
is through {}``addition'' that the principle of conservation of
momentum is expressed. We will look into this in more detail in Section
3, but in its essence the addition rule is a map
\begin{eqnarray}
\oplus:\,\mathcal{P}\times\mathcal{P} & \rightarrow & \mathcal{P}\nonumber \\
(p,q) & \rightarrow & p\oplus q
\end{eqnarray}
which can be experimentally determined from particle collision processes,
where total momentum is conserved. For very low energies, it should
reduce to the usual linear addition, $p\oplus q\approx p+q$, but
in the spirit of DSR, we will assume that in general it is not linear.
But if $\oplus$ is not linear, there is no reason to assume it is
commutative or even associative - and we will not do so. However,
to turn incoming momenta into outgoing and vice-versa while analyzing
point particle interactions, we will require an inversion map, i.e.
to assume that each momentum $p$ has a reciprocal $\ominus p$ satisfying
both the left and right inverse conditions:
\begin{equation}
(\ominus p)\oplus(p\oplus q)=(q\oplus p)\oplus(\ominus p)=q,\,\,\,\,\forall p,q\in\mathcal{P}
\end{equation}
$\oplus$ is directly related to the notion of a connection in $\mathcal{P}$,
since it can be used to define parallel transport of a vector. As
shown in Figure 2, the parallel transport of the vector field $p$
from $\mathcal{T}_{0}\mathcal{P}$ to $\mathcal{T}_{dq}\mathcal{P}$
results in their combination $p\oplus dq$, which admits the following
expansion for small $p$ assuming $dq$ is infinitesimal:
\begin{equation}
\left(p\oplus dq\right)_{a}=p_{a}+dq_{a}-\Gamma_{a}^{bc}(0)\, p_{b}dq_{c}+(...)
\end{equation}
\begin{figure}[H]
\begin{centering}
\includegraphics{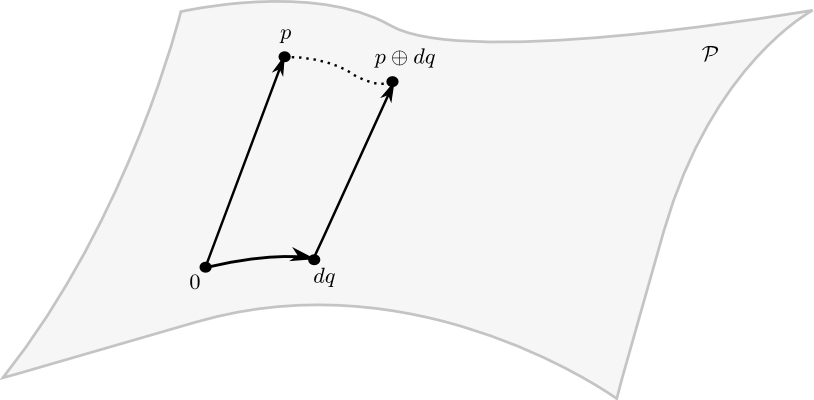}
\par\end{centering}

\caption{Momentum addition as a parallel transport.}
\end{figure}

The definition of the connection of momentum space is then natural:
\begin{equation}
\Gamma_{a}^{bc}(k)=-\left.\frac{\partial}{\partial p_{b}}\frac{\partial}{\partial q_{c}}(p\oplus_{k}q)_{a}\right|_{p=q=k}\label{eq:connection}
\end{equation}
where $\oplus_{k}$ is simply the addition rule with the neutral element
shifted from 0 to $k$, which has the expression
\begin{equation}
p\oplus_{k}q=k\oplus((\ominus k\oplus p)\oplus(\ominus k\oplus q))
\end{equation}
and satisfies $k\oplus_{k}k=k,\,\forall k\in\mathcal{P}$ (analogous
to $0\oplus0=0$).\\
\\
One important property of our constructions of the metric and connection
of momentum space is that they are independent of one another - they
result from different measurements. This means that, in general, the
connection and the metric might not be compatible. We can quantify
the incompatibility using the \textit{non-metricity tensor}, $N^{abc}=\nabla^{a}g^{bc}.$
It is possible to decompose the full connection in terms of the Levi-Civita
connection, the torsion and the non-metricity. To do this, decompose
the connection components in their symmetric and antisymmetric parts
with respect to the top indices:
\begin{equation}
\Gamma_{c}^{ab}=\Gamma_{c}^{(ab)}+\Gamma_{c}^{[ab]}=\left\{ _{\,\, c}^{ab}\right\} +\frac{1}{2}T_{c}^{ab}+\mathcal{N}_{c}^{ab}
\end{equation}
where $\mathcal{N}_{c}^{ab}=\Gamma_{c}^{(ab)}-\left\{ _{\,\, c}^{ab}\right\} $
is symmetric in the top indices. Now it can be shown using the definition
of covariant derivative that (defining the covariant torsion $T^{abc}=T_{d}^{ab}g^{dc}$)
\begin{eqnarray}
N^{abc} & = & \frac{1}{2}\left(T^{abc}+T^{acb}\right)-\mathcal{N}_{d}^{ba}g^{dc}-\mathcal{N}_{d}^{ca}g^{db}\\
\Rightarrow N^{abi}+N^{bai}-N^{iab} & = & -T^{iab}-T^{iba}-2\mathcal{N}_{d}^{ab}g^{di}
\end{eqnarray}
so that the connection can be written as
\begin{equation}
\Gamma_{c}^{ab}=\left\{ _{\,\, c}^{ab}\right\} +\frac{1}{2}T_{c}^{ab}-\frac{1}{2}g_{ci}\left(N^{abi}+N^{bai}-N^{iab}+T^{iab}+T^{iba}\right)\label{eq:decomp}
\end{equation}

\subsubsection*{Curvature of momentum space}

From the connection introduced above, we can construct the Riemann
tensor of momentum space. The definition in terms of the combination
rule is
\begin{equation}
R_{\quad d}^{abc}(r)=\left.2\frac{\partial}{\partial p_{[a}}\frac{\partial}{\partial q_{b]}}\frac{\partial}{\partial k_{c}}\left(\left[\left(p\oplus_{r}q\right)\oplus_{r}k\right]-\left[p\oplus_{r}\left(q\oplus_{r}k\right)\right]\right)_{d}\right|_{p=q=k=r}\label{eq:riemann}
\end{equation}
While the connection describes the non-linearity of $\oplus$, the
curvature is a measure of its non-associativity.

\begin{onehalfspace}

\section{Dynamics of point particle interactions}
\end{onehalfspace}

\begin{onehalfspace}

\subsection{General considerations%
\footnote{Review of content described in \cite{key-1},\cite{key-2}.%
}}
\end{onehalfspace}

Consider a system of \textit{N} point particles labeled by indices
$I$ interacting locally at vertices $\alpha$, with worldlines $k_{a}^{I}(s)\in\mathcal{P},\, s\in\mathcal{I}\subset\mathbb{R}$.
This system is described by the action \cite{key-1}
\begin{eqnarray}
S & = & \sum_{I}S_{free}+\sum_{\alpha}S_{int}\nonumber \\
 & = & \sum_{I}\int ds\left(x_{I}^{a}\dot{k}_{a}^{I}+\mathcal{N}_{I}C^{I}(k^{I})\right)+\sum_{\alpha}\left(-\mathcal{K}_{a}^{(\alpha)}z_{(\alpha)}^{a}\right).
\end{eqnarray}
In this expression, we introduce spacetime coordinates $x_{I}^{a}$
as elements of $\mathcal{T}_{k}^{*}\mathcal{P}$ canonically dual
to the momenta, satisfying the Poisson brackets $\left\{ x^{a},\, k_{b}\right\} =\delta_{b}^{a}$.
The metric of spacetime plays no role though; only the metric of momentum
space appears in the action through the constraints
\begin{equation}
C^{I}(k)=D^{2}(k)-m_{I}^{2},
\end{equation}
which specify the particle masses. The interaction term imposes conservation
of momentum in each vertex $\alpha$ through the constraints $\mathcal{K}_{a}^{(\alpha)}=\mathcal{K}_{a}^{(\alpha)}\left(k_{(\alpha)}^{I}\right)$,
which depend on the vertex endpoints of the interacting worldlines
and are related via equations of motion to the interaction coordinates
$z_{(\alpha)}^{a}$ in spacetime.

\begin{figure}[H]
\begin{centering}
\includegraphics{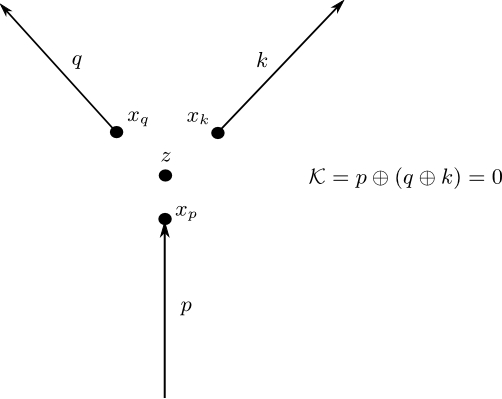}
\par\end{centering}

\caption{A typical interaction vertex.}
\end{figure}

The free equations of motion for this action,
\begin{eqnarray}
\frac{\delta S}{\delta k_{a}^{I}} & = & 0\Rightarrow\dot{x}_{I}^{a}=\mathcal{N}_{I}\frac{\delta C^{I}}{\delta k_{a}^{I}}\\
\frac{\delta S}{\delta x_{I}^{a}} & = & 0\Rightarrow\dot{k}_{a}^{I}=0\\
\frac{\delta S}{\delta\mathcal{N}_{I}} & = & 0\Rightarrow C^{I}(k)=0,
\end{eqnarray}
are analogous to the usual spacetime-centered perspective in classical
mechanics. There are two more equations to be derived though - we
can still vary the $z_{(\alpha)}$ and the $k_{(\alpha)}^{I}$ at
the interaction vertices.
\begin{itemize}
\item Varying with respect to $z_{(\alpha)}^{a}$ results in the momentum
conservation constraints in each vertex, given by
\begin{equation}
\mathcal{K}_{a}^{(\alpha)}=0;
\end{equation}

\item varying with respect to $k_{(\alpha)}^{I}$ results in the \textit{relative
locality relations} between the location in spacetime of the worldline
endpoints and the vertex coordinates:
\begin{equation}
\boxed{x_{I,\alpha}^{a}=z_{(\alpha)}^{b}\left(\pm\frac{\delta\mathcal{K}_{b}^{(\alpha)}}{\delta k_{a}^{I,\alpha}}\right)\equiv z_{(\alpha)}^{b}\left(W_{x_{I}}^{(\alpha)}\right)_{b}^{a}}
\end{equation}
where the $+$ sign refers to incoming particles and $-$ to outgoing
ones. Note that this variation requires paying attention to whether
the corresponding worldline is incoming or outgoing with respect to
the vertex, i.e. assuming for simplicity that $\mathcal{I}=[0,1]$,
$k_{(\alpha)}^{I}=k^{I}(1)$ or $k^{I}(0)$ if the particle is incoming
or outgoing respectively.
\end{itemize}
From the relative locality relations we see that not only do the worldline
endpoints in spacetime not coincide with the vertex location, but,
since the operators $W_{x_{I}}$ are related to parallel transport
(as we will see in more detail later), they do so in a way that directly
reflects the fact that the different particles are in {}``different
spacetimes''. For very low energies (small momenta), we recover the
usual notion of locality, $\left(W_{x_{I}}\right)_{b}^{a}\approx\delta_{b}^{a}\Rightarrow x_{I,\alpha}^{a}=z_{(\alpha)}^{a}$.\\
\\
For the following, we will require some definitions and two useful
results.

\subsubsection*{Translation and parallel transport operators}

The aforementioned addition rule for momenta defines left- and right-
translation operators:
\begin{equation}
p\oplus q\equiv L_{p}(q)\equiv R_{q}(p),\,\,\,\,\forall p,q\in\mathcal{P}
\end{equation}
Based on our interpretation of parallel transport through addition
described in section 2.2, it is natural to define parallel transport
operators as derivatives of the translation operators:
\begin{equation}
\left(U_{p\oplus q}^{q}\right)_{a}^{b}\equiv\frac{\partial(p\oplus q)_{a}}{\partial q_{b}}=(d_{q}L_{p})_{a}^{b},\label{eq:U}
\end{equation}
\begin{equation}
\left(V_{p\oplus q}^{p}\right)_{a}^{b}\equiv\frac{\partial(p\oplus q)_{a}}{\partial p_{b}}=(d_{p}R_{q})_{a}^{b}.\label{eq:V}
\end{equation}
We define an additional operator as the derivative of the inversion
map $\ominus$: 
\begin{equation}
\left(I^{p}\right)_{a}^{b}\equiv\frac{\partial(\ominus p)_{a}}{\partial p_{b}}=(d_{p}\ominus)_{a}^{b}.\label{eq:I}
\end{equation}
Note that $L,\, R$ are diffeomorphisms in $\mathcal{P}$, while $U,\, V,\, I$
are linear operators in $T\mathcal{P}$, and that the usual chain
rule of differentiation applies, $d_{p}(A\circ B)=d_{B(p)}A\circ d_{p}B,\,\,\forall A,\, B\in\mathcal{E}(\mathcal{P})$%
\footnote{$\mathcal{E}(\mathcal{P})\equiv$ diffeomorphisms in $\mathcal{P}$%
}. We have the following results \cite{key-2}:
\begin{enumerate}
\item Inverses of \textit{U}, \textit{V}, \textit{I}:
\begin{equation}
\left(U_{q}^{p}\right)^{-1}=U_{p}^{q};\qquad\left(V_{q}^{p}\right)^{-1}=V_{p}^{q};\qquad\left(I^{p}\right)^{-1}=I^{\ominus p}.
\end{equation}
To see this, start by writing $U_{q}^{p}=U_{(q\ominus p)\oplus p}^{p}=d_{p}L_{q\ominus p}$
and $U_{p}^{q}=U_{(p\ominus q)\oplus q}^{q}=d_{q}L_{p\ominus q}$.
Then it is clear that%
\footnote{We use the property $\ominus(p\oplus q)=\ominus q\ominus p$. Proof
is as follows: $\ominus(p\oplus q)=\ominus q\ominus p\Leftrightarrow q\ominus(p\oplus q)=\ominus p\Leftrightarrow(q\ominus(p\oplus q))\oplus(p\oplus q)=\ominus p\oplus(p\oplus q)\Leftrightarrow q=q$.%
} $\mathbf{1}=d_{p}(Id)=d_{p}L_{(p\ominus q)\oplus(q\ominus p)}=d_{p}\left(L_{p\ominus q}L_{q\ominus p}\right)=d_{q}L_{p\ominus q}d_{p}L_{q\ominus p}=U_{p}^{q}U_{q}^{p}$.
The calculation for \textit{V} is entirely analogous. For \textit{I}
note that%
\footnote{Note that $\ominus:\,\mathcal{P}\rightarrow\mathcal{P},\, p\rightarrow\ominus p\in\mathcal{E}(\mathcal{P})$
is its own inverse of since $\ominus(\ominus p)=p.$%
} $\mathbf{1}=d_{p}(Id)=d_{p}(\ominus\ominus)=d_{\ominus p}\ominus\cdot d_{p}\ominus=I^{\ominus p}I^{p}$.
\item Relation between \textit{U}, \textit{V}, \textit{I}:
\begin{equation}
-U_{0}^{\ominus p}I^{p}=V_{0}^{p}.\label{eq:result2}
\end{equation}
The fact that $(\ominus p)\oplus(p\oplus q)=q$ implies that $d_{p}\left((\ominus p)\oplus(p\oplus q)\right)=0$.
We can express this in terms of the parallel transport operators as
\begin{eqnarray}
0=d_{p}\left((\ominus p)\oplus(p\oplus q)\right) & = & \left.\left[d_{p}\left((\ominus p')\oplus(p\oplus q)\right)+d_{p}\left((\ominus p)\oplus(p'\oplus q)\right)\right]\right|_{p=p'}\nonumber \\
 & = & \left.\left[d_{p}\left(L_{\ominus p'}R_{q}\right)+d_{p}\left(R_{p'\oplus q}\ominus\right)\right]\right|_{p=p'}\nonumber \\
 & = & d_{R_{q}p}L_{\ominus p}\cdot d_{p}R_{q}+d_{\ominus p}R_{p\oplus q}\cdot d_{p}\ominus\nonumber \\
 & = & U_{q}^{p\oplus q}V_{p\oplus q}^{p}+V_{q}^{\ominus p}I^{p}.
\end{eqnarray}
Fixing $q=0$, we then get
\begin{equation}
U_{0}^{p}V_{p}^{p}+V_{0}^{\ominus p}I^{p}=0.
\end{equation}
In point 1 we showed that $V_{p}^{p}=\boldsymbol{1}$ and $\left(I^{p}\right)^{-1}=I^{\ominus p}$,
so applying $I^{\ominus p}$ to the equation and renaming $p\rightarrow\ominus p$,
we get the final result.
\end{enumerate}

\subsubsection*{The bivalent vertex}

To better understand the relative locality formalism, we will consider
the simplest example of an {}``interaction'' vertex: propagation
of a free particle. There are two momenta in the corresponding diagram
- one incoming and one outgoing, as Figure 4 illustrates:

\begin{figure}[H]
\begin{centering}
\includegraphics{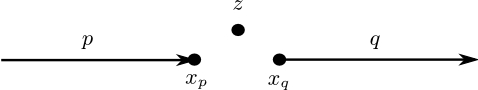}
\par\end{centering}

\caption{The bivalent vertex.}
\end{figure}

Physically, we do not expect relative locality to introduce any effects
in this situation, since any worldline of a free particle can be broken
into segments separated by bivalent vertices, introducing momentum
constraints in each one - and we do not expect the different worldline
segments created by this process to be in different spacetimes. Let
us verify this in practice by computing the relative locality relations.
Write the conservation rule as
\begin{equation}
\mathcal{K}=p\ominus q=0.
\end{equation}
The relative locality operators are
\begin{eqnarray}
W_{x_{p}} & = & d_{p}\mathcal{K}=d_{p}R_{\ominus q}=V_{p\ominus q}^{p}=V_{0}^{p}=V_{0}^{q},\\
W_{x_{q}} & = & -d_{q}\mathcal{K}=-d_{q}\left(L_{p}\ominus\right)=-d_{\ominus q}L_{p}\cdot d_{q}\ominus=-U_{p\ominus q}^{\ominus q}I^{q}=-U_{0}^{\ominus q}I^{q},
\end{eqnarray}
where we used the momentum constraint to simplify results. Using \eqref{eq:result2},
we get $W_{x_{q}}=W_{x_{p}}$, so that the worldline endpoint coordinates
are
\begin{equation}
x_{p}^{a}=x_{q}^{a}=z^{b}\left(V_{0}^{q}\right)_{b}^{a},
\end{equation}
and the {}``interaction'' is local to every observer - the only
effect of a coordinate change is to alter its location with respect
to the chosen frame, as one would expect in ordinary Special Relativity.\\
\\
Much more interesting is the case of the trivalent vertex, which can
represent physical situations such as the decay of one particle into
two or the fusion of two particles. This is the simplest situation
where we expect relative locality to introduce novel effects, and
it will be the primary kind of vertex studied in the following sections.

\subsection{The change of vertex problem }

A generic momentum constraint $\mathcal{K}$ can naively be written
as
\begin{equation}
\mathcal{K}_{b}=\overset{N}{\underset{J=1}{\oplus}}k_{b}^{J}\label{eq:naivemomentumconstraint}
\end{equation}
generalizing the usual linear conservation laws. However, because
the addition rule is in general non-commutative and non-associative,
\eqref{eq:naivemomentumconstraint} is ambiguous: while in the linear
case the ordering of additions is naturally irrelevant, in this more
general setting it changes the form of the constraint. It is then
important to evaluate if these changes affect the physics of the interaction,
and in the event they do, determine a consistent, physical way of
choosing vertex orderings. A bivalent vertex has two possible ways
of writing the momentum constraint, $\mathcal{K}=p\ominus q$ or $\mathcal{K}=q\ominus p$,
while a trivalent vertex has 12, corresponding to all the different
permutations and operation orderings of the 3 momenta (if they are
all regarded as incoming for simplicity, some of the possibilities
are $\mathcal{K}^{(1)}=(p\oplus q)\oplus k$, $\mathcal{K}^{(2)}=p\oplus(q\oplus k)$,
$\mathcal{K}^{(3)}=p\oplus(k\oplus q)$...).\\
\\
Notice that changing the form of writing a single momentum constraint
corresponds effectively to applying a diffeomorphism $\lambda$ on
the original one. For example, $\mathcal{K}^{(2)}=R_{q\oplus k}R_{\ominus q}R_{\ominus k}\mathcal{K}^{(1)}$
and $\mathcal{K}^{(3)}=R_{k\oplus q}R_{\ominus k\ominus q}\mathcal{K}^{(2)}$.
The diffeomorphism that relates two different constraints is not unique
- we could also have written $\mathcal{K}^{(2)}=L_{p}L_{q}L_{\ominus q\ominus p}\mathcal{K}^{(1)}$.
If we proceed under the assumption there should be at least a class
of vertex changes that does not affect the physics of the interaction,
we should try to find something analogous to a \textit{gauge transformation}
on our system based on the above considerations, i.e. a redundancy
in the definition of momentum constraints and vertex coordinates that
does not affect the equations of motion. Indeed, there is such a transformation,
as described originally in \cite{key-7}:
\begin{equation}
\mathcal{K}_{a}^{(\alpha)}\rightarrow\lambda_{a}^{(\alpha)}\left(\mathcal{K}^{(\alpha)}\right);\qquad z_{(\alpha)}^{a}\rightarrow z_{(\alpha)}^{b}\frac{\partial\mathcal{K}_{b}^{(\alpha)}}{\partial\lambda_{a}^{(\alpha)}}\qquad\lambda^{(\alpha)}\in\mathcal{E}(\mathcal{P}),\,\lambda^{(\alpha)}(0)=0
\end{equation}
This rewriting neither alters the constraints, since $\mathcal{K}^{(\alpha)}=0\Leftrightarrow\lambda^{(\alpha)}\left(\mathcal{K}^{(\alpha)}\right)=0$,
nor affects the worldline endpoint coordinates given by the relative
locality relations:
\begin{eqnarray}
x_{I,\alpha}^{a} & = & z_{(\alpha)}^{b}\left(\pm\frac{\delta\mathcal{K}_{b}^{(\alpha)}}{\delta k_{a}^{I,\alpha}}\right)\;\rightarrow\; z_{(\alpha)}^{c}\frac{\partial\mathcal{K}_{c}^{(\alpha)}}{\partial\lambda_{b}^{(\alpha)}}\cdot\left(\pm\frac{\delta\lambda_{b}^{(\alpha)}\left(\mathcal{K}^{(\alpha)}\right)}{\delta k_{a}^{I,\alpha}}\right)\nonumber \\
 & = & z_{(\alpha)}^{c}\frac{\partial\mathcal{K}_{c}^{(\alpha)}}{\partial\lambda_{b}^{(\alpha)}}\cdot\left(\pm\frac{\partial\lambda_{b}^{(\alpha)}}{\partial\mathcal{K}_{d}^{(\alpha)}}\frac{\delta\mathcal{K}_{d}^{(\alpha)}}{\delta k_{a}^{I,\alpha}}\right)\nonumber \\
 & = & z_{(\alpha)}^{c}\left(\pm\frac{\delta\mathcal{K}_{c}^{(\alpha)}}{\delta k_{a}^{I,\alpha}}\right)=x_{I,\alpha}^{a}.
\end{eqnarray}
It is clear that the physics of an interaction is unaffected by the
map above, which means we can regard it as a gauge symmetry of the
theory.\\
\\
As an application of this idea, consider once again the bivalent vertex.
In Section 3.1 we used the momentum constraint $\mathcal{K}^{(1)}=p\ominus q$.
Let us see what happens if we use $\mathcal{K}^{(2)}=q\ominus p$.
The relative locality operators are then
\begin{eqnarray}
W_{x_{p}}^{(2)} & = & U_{0}^{\ominus p}I^{p}=-V_{0}^{p}=-V_{0}^{q}\quad\text{and}\quad W_{x_{q}}^{(2)}=-V_{0}^{q},
\end{eqnarray}
so that the worldline endpoint coordinates are given by 
\begin{eqnarray}
x_{p}^{a} & = & x_{q}^{a}=\left(-z^{b}\right)\left(V_{0}^{q}\right)_{b}^{a}\equiv z'^{b}\left(V_{0}^{q}\right)_{b}^{a}.
\end{eqnarray}
This can be described as the result of a gauge transformation if we
note that $\mathcal{K}^{(2)}=R_{\ominus p}\cdot\ominus\cdot L_{\ominus p}\mathcal{K}^{(1)}\equiv\lambda\left(\mathcal{K}^{(1)}\right)$
and calculate
\begin{eqnarray}
z' & = & z\frac{\partial\mathcal{K}^{(1)}}{\partial\lambda}=z\left(\frac{\partial\lambda}{\partial\mathcal{K}^{(1)}}\right)^{-1}=z\,\left[d_{0}\left(R_{\ominus p}\cdot\ominus\cdot L_{\ominus p}\right)\right]^{-1}\nonumber \\
 & = & z\left(V_{0}^{p}I^{\ominus p}U_{\ominus p}^{0}\right)^{-1}=-z\left(U_{0}^{\ominus p}U_{\ominus p}^{0}\right)^{-1}\nonumber \\
 & = & -z.
\end{eqnarray}
We have seen a simple example of how a certain class of vertex changes
do not affect the interaction physics, since it can actually be matched
to a redundancy in the definition of the vertex coordinates together
with the conservation laws. However, notice that not all diffeomorphisms
corresponding to vertex changes obey the condition $\lambda(0)=0$:
an example was given above, $R_{k\oplus q}R_{\ominus k\ominus q}(0)=(\ominus k\ominus q)\oplus(k\oplus q)\neq0$
unless $\oplus$ is commutative. These cases motivate further study,
and some concrete examples will be given in the next section.

\begin{onehalfspace}

\subsection{Illustration: gamma ray bursts}
\end{onehalfspace}

A typical gamma ray burst event consists of a high-energy gamma ray
emission followed by an {}``afterglow'' of lower energy radiation
(typically X-rays)\cite{key-13}. It has been argued in \cite{key-2}
that observational measurements of phenomena of this kind can serve
as a detector of non-metricity and/or torsion of momentum space, via
the time delay between the detection of two photons of different energies
emitted by the same source and the angle deflection relative to the
direction of the sources (dual gravitational lensing), respectively.
We will redo the calculation of \cite{key-2} in a slightly different
way, to make transparent where a change of vertex could alter the
form of the results and/or the physics, and we will consider three
possible changes in the form of the momentum constraints and reflect
on what the similarities and differences between the calculations
in each one are.\\
\\
The idealized description of the gamma ray burst is summarized in
Figure 5. The {}``experimental set-up'' consists of an emitter and
a detector, represented by particles with masses $m_{1}\text{ and }m_{2}$
respectively. The emitter sends a photon with energy $E_{1}$, and,
after a proper time $S_{1}$, it sends another photon with energy
$E_{2}\neq E_{1}$. The light rays take proper times $T_{1}$ and
$T_{2}$ respectively to reach the detector, and the proper time interval
between the two detections is $S_{2}$.

\begin{figure}[H]
\noindent \begin{centering}
\includegraphics{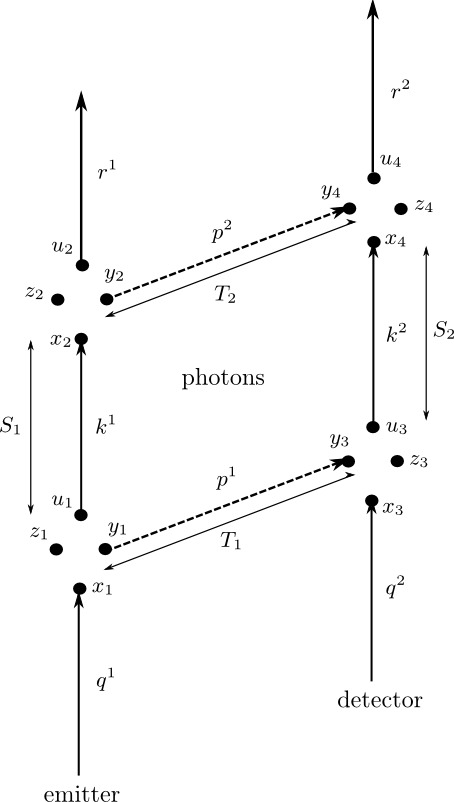}
\par\end{centering}

\caption{Notation of positions and momenta in the gamma ray burst setup \cite{key-2}.}
\end{figure}

Special Relativity predicts that, assuming the emitter and the detector
are at rest with respect to each other, $\frac{k^{1}}{m_{1}}=\frac{k^{2}}{m_{2}}$,
the relation between the intervals of emission and detection is the
intuitive one, $S_{1}=S_{2}$. However, working in the relative locality
formalism it is possible to have a time delay between the two events,
depending on the energies of the photons sent, as we will see.\\
\\
The derivation of the time delay $\Delta S=S_{2}-S_{1}$ begins with
noting that spacetime metric is Minkowski (since we are still working
on the $G_{N}\rightarrow0,\,\hbar\rightarrow0$ limit), so the equation
of motion for the spacetime worldlines is simply (picking the $\mathcal{N}_{I}$
appropriately) $\dot{x}_{I}^{a}=g^{ab}k_{b}^{I}\equiv k_{I}^{a}$,
and it is then possible to establish kinematic relations between the
momenta and the proper times of photon propagation, which are the
invariant quantities that we need to calculate $\Delta S$:
\begin{eqnarray}
x_{2}-u_{1} & = & \hat{k}_{1}S_{1};\qquad x_{4}-u_{3}=\hat{k}_{2}S_{2}\nonumber \\
y_{3}-y_{1} & = & \hat{p}_{1}T_{1};\qquad y_{4}-y_{2}=\hat{p}_{2}T_{2}\label{eq:kin_id}
\end{eqnarray}
where $\hat{k}_{i}=\frac{k_{i}}{m_{i}}$ and $\hat{p}_{i}=\frac{p_{i}}{E_{i}}$
,$\forall i$.\\
\\
The next step is to use \eqref{eq:kin_id} to construct a series of
identities relating the proper times $S_{i},T_{i}$ by taking two
different routes from $z_{1}$ to $z_{4}$, forming a loop in spacetime.
We would like to eliminate the interaction coordinates from these
identities, so that the expression obtained only includes diffeomorphism-covariant
quantities. We use the relative locality relations $x_{I,\alpha}=z_{(\alpha)}W_{x_{I}}^{(\alpha)}$
to obtain 
\begin{equation}
T_{1}\hat{p}_{1}W_{y_{3}}^{-1}=(y_{3}-y_{1})W_{y_{3}}^{-1}=z_{3}-z_{1}W_{y_{1}}W_{y_{3}}^{-1}
\end{equation}
\begin{equation}
S_{2}\hat{k}_{2}W_{u_{3}}^{-1}=(x_{4}-u_{3})W_{u_{3}}^{-1}=z_{4}W_{x_{4}}W_{u_{3}}^{-1}-z_{3}
\end{equation}
Eliminate $z_{3}$ from these to get
\begin{equation}
T_{1}\hat{p}_{1}W_{y_{3}}^{-1}+S_{2}\hat{k}_{2}W_{u_{3}}^{-1}=z_{4}W_{x_{4}}W_{u_{3}}^{-1}-z_{1}W_{y_{1}}W_{y_{3}}^{-1}\label{eq:z4z1a}
\end{equation}
Similarly, eliminating $z_{2}$ from the identities
\begin{eqnarray}
T_{2}\hat{p}_{2}W_{y_{4}}^{-1} & = & (y_{4}-y_{2})W_{y_{4}}^{-1}=z_{4}-z_{2}W_{y_{2}}W_{y_{4}}^{-1}\Leftrightarrow T_{2}\hat{p}_{2}W_{y_{2}}^{-1}=z_{4}W_{y_{4}}W_{y_{2}}^{-1}-z_{2}
\end{eqnarray}
\begin{equation}
S_{1}\hat{k}_{1}W_{x_{2}}^{-1}=z_{2}-z_{1}W_{u_{1}}W_{x_{2}}^{-1}
\end{equation}
we get
\begin{equation}
S_{1}\hat{k}_{1}W_{x_{2}}^{-1}+T_{2}\hat{p}_{2}W_{y_{2}}^{-1}=z_{4}W_{y_{4}}W_{y_{2}}^{-1}-z_{1}W_{u_{1}}W_{x_{2}}^{-1}\label{eq:z4z1b}
\end{equation}
\eqref{eq:z4z1a} and \eqref{eq:z4z1b} involve only $z_{4}$ and
$z_{1}$. Eliminate $z_{4}$ to obtain
\begin{eqnarray}
z_{1}\left(W_{u_{1}}W_{x_{2}}^{-1}W_{y_{2}}W_{y_{4}}^{-1}-W_{y_{1}}W_{y_{3}}^{-1}W_{u_{3}}W_{x_{4}}^{-1}\right) & = & S_{2}K_{2}-S_{1}K_{1}-T_{2}P_{2}+T_{1}P_{1}\label{eq:tdelay}
\end{eqnarray}
where we defined the auxiliary momenta
\begin{eqnarray}
K_{1} & \equiv & \hat{k}_{1}W_{x_{2}}^{-1}W_{y_{2}}W_{y_{4}}^{-1}\nonumber \\
K_{2} & \equiv & \hat{k}_{2}W_{x_{4}}^{-1}\nonumber \\
P_{1} & \equiv & \hat{p}_{1}W_{y_{3}}^{-1}W_{u_{3}}W_{x_{4}}^{-1}\nonumber \\
P_{2} & \equiv & \hat{p}_{2}W_{y_{4}}^{-1}\label{eq:uppercase momenta}
\end{eqnarray}
\eqref{eq:tdelay} is the key equation that will allow us to derive
the time delay, since if its LHS vanishes, we have successfully derived
a relation between the invariant proper times and diffeomorphism-covariant
momenta.\\
\\
We will now specify momentum constraints and calculate the relevant
quantities for our study for each of our choices. For simplicity,
we will only do computations to first order in the momenta. Definitions
\eqref{eq:connection}, \eqref{eq:U}, \eqref{eq:V} and \eqref{eq:I}
allow us to expand the parallel transport operators:
\begin{eqnarray}
\left(U_{p_{2}}^{p_{1}}\right)_{a}^{b} & = & \delta_{a}^{b}-\Gamma_{a}^{cb}(p_{2}-p_{1})_{c}+\mathcal{O}(p_{2}-p_{1})^{2}\nonumber \\
\left(V_{p_{2}}^{p_{1}}\right)_{a}^{b} & = & \delta_{a}^{b}-\Gamma_{a}^{bc}(p_{2}-p_{1})_{c}+\mathcal{O}(p_{2}-p_{1})^{2}\nonumber \\
\left(I^{p}\right)_{a}^{b} & = & -\delta_{a}^{b}-(\Gamma_{a}^{bc}+\Gamma_{a}^{cb})p_{c}+\mathcal{O}(p^{2})\label{eq:parl-trans-exp}
\end{eqnarray}

\subsubsection{Specifying momentum constraints}

The choice of conservation laws in \cite{key-2} is
\begin{equation}
\begin{array}{cccccc}
\mathcal{K}^{1}= & (q^{1}\ominus k^{1})\ominus p^{1} & =0\qquad & \mathcal{K}^{2}= & (k^{1}\ominus r^{1})\ominus p^{2} & =0\\
\mathcal{K}^{3}= & p^{1}\oplus(\ominus k^{2}\oplus q^{2}) & =0\qquad & \mathcal{K}^{4}= & p^{2}\oplus(\ominus r^{2}\oplus k^{2}) & =0
\end{array}
\end{equation}
from which we can compute the relative locality operators:
\begin{equation}
\begin{array}{cccccccc}
W_{x_{1}}= & V_{0}^{p^{1}}V_{p^{1}}^{q^{1}} & \qquad & W_{u_{1}}= & -V_{0}^{p^{1}}U_{p^{1}}^{\ominus k^{1}}I^{k^{1}} & \qquad & W_{y_{1}}= & V_{0}^{p^{1}}\\
W_{x_{2}}= & V_{0}^{p^{2}}V_{p^{2}}^{k^{1}} & \qquad & W_{u_{2}}= & -V_{0}^{p^{2}}U_{p^{2}}^{\ominus r^{1}}I^{r^{1}} & \qquad & W_{y_{2}}= & V_{0}^{p^{2}}\\
W_{x_{3}}= & U_{0}^{\ominus p^{1}}U_{\ominus p^{1}}^{q^{2}} & \qquad & W_{u_{3}}= & -U_{0}^{\ominus p^{1}}V_{\ominus p^{1}}^{\ominus k^{2}}I^{k^{2}} & \qquad & W_{y_{3}}= & V_{0}^{p^{1}}\\
W_{x_{4}}= & U_{0}^{\ominus p^{2}}U_{\ominus p^{2}}^{k^{2}} & \qquad & W_{u_{4}}= & -U_{0}^{\ominus p^{2}}V_{\ominus p^{2}}^{\ominus r^{2}}I^{r^{2}} & \qquad & W_{y_{4}}= & V_{0}^{p^{2}}
\end{array}
\end{equation}
Performing the 1st order expansions of these operators using \eqref{eq:parl-trans-exp}
and computing the quantities in \eqref{eq:tdelay}, it was found in
\cite{key-2} that
\begin{itemize}
\item $\left(W_{u_{1}}W_{x_{2}}^{-1}W_{y_{2}}W_{y_{4}}^{-1}\right)_{a}^{b}\approx\delta_{a}^{b}+T_{a}^{bc}p_{c}^{1}\approx W_{y_{1}}W_{y_{3}}^{-1}W_{u_{3}}W_{x_{4}}^{-1}$,
so that the LHS of \eqref{eq:tdelay} conveniently vanishes to 1st
order;
\item The momenta \eqref{eq:uppercase momenta} have the approximate expressions
\begin{eqnarray}
K_{1}^{b} & \approx & \hat{k}_{1}^{b}-m_{1}\hat{k}_{1}^{a}\Gamma_{a}^{bc}\hat{k}_{c}^{1}\approx\left(\hat{k}_{1}V_{k^{1}}^{0}\right)^{b}\nonumber \\
K_{2}^{b} & \approx & \hat{k}_{2}^{b}-m_{2}\hat{k}_{2}^{a}\Gamma_{a}^{cb}\hat{k}_{c}^{2}\approx\left(\hat{k}_{2}U_{k^{2}}^{0}\right)^{b}\nonumber \\
P_{1}^{b} & \approx & \hat{p}_{1}^{b}-E_{1}\hat{p}_{1}^{a}\Gamma_{a}^{cb}\hat{p}_{c}^{1}\approx\left(\hat{p}_{1}U_{p^{1}}^{0}\right)^{b}\nonumber \\
P_{2}^{b} & \approx & \hat{p}_{2}^{b}-E_{2}\hat{p}_{2}^{a}\Gamma_{a}^{bc}\hat{p}_{c}^{2}\approx\left(\hat{p}_{2}V_{p^{2}}^{0}\right)^{b}\label{eq:origmomenta}
\end{eqnarray}
where we observe that each uppercase momentum depends only on the
corresponding lowercase momentum. The approximate expressions in terms
of $U\text{ and }V$ mean that the uppercase momenta can be interpreted
as parallel transports of the lowercase ones from $T_{p_{i},k_{i}}^{*}\mathcal{P}$
back to $T_{0}^{*}\mathcal{P}$.
\end{itemize}
The paper \cite{key-2} goes on to relate the existence of non-metricity
to a nonzero time delay through the first order version of \eqref{eq:tdelay},
$K_{2}S_{2}-K_{1}S_{1}=P_{2}T_{2}-P_{1}T_{1}$. Working in Riemann
normal coordinates at the origin of momentum space, so that the metric
near 0 is $g^{ab}(k)\approx\eta^{ab},\,\left|k\right|\ll1$, and supposing
that the torsion vanishes, the connection is entirely given by the
non-metricity portion in \eqref{eq:decomp} and the norms of the uppercase
momenta can be calculated:
\begin{equation}
K_{i}^{2}\approx-1+m_{i}N^{abc}\hat{k}_{a}^{i}\hat{k}_{b}^{i}\hat{k}_{c}^{i};\qquad P_{i}^{2}\approx E_{i}N^{abc}\hat{p}_{a}^{i}\hat{p}_{b}^{i}\hat{p}_{c}^{i}\label{eq:nonmetricnorms}
\end{equation}
The assumption analogous to $\hat{k}_{1}=\hat{k}_{2}$ in the relative
locality setup is to make the parallel transported $\hat{K_{i}}$
vectors parallel to each other - define $\hat{K}=\hat{K}_{1}=\hat{K}_{2}$.
\eqref{eq:tdelay} then resolves to
\begin{equation}
\hat{K}\Delta S\equiv\hat{K}\left(\left|K_{2}\right|S_{2}-\left|K_{1}\right|S_{1}\right)=T_{2}P_{2}-T_{1}P_{1}\label{eq:tdelaynew}
\end{equation}
We can now decompose the photon momenta in a component parallel to
$\hat{K}$ and a component parallel to unit spacelike vectors $R_{i}\perp\hat{K}$,
\begin{equation}
P_{i}=\left(\hat{K}\cdot P_{i}\right)\hat{K}+\sqrt{\left(\hat{K}\cdot P_{i}\right)^{2}-P_{i}^{2}}\, R_{i}.\label{eq:photondecomp}
\end{equation}
Notice that taking the scalar product of \eqref{eq:tdelaynew} with
$\hat{K}$ we get the formula for the time delay, 
\begin{equation}
\Delta S=\left(\hat{K}\cdot P_{2}\right)T_{2}-\left(\hat{K}\cdot P_{1}\right)T_{1},\label{eq:tdelayscalar1}
\end{equation}
and taking the scalar product of \eqref{eq:tdelaynew} with $\hat{R}_{i}$
we get $\left(T_{2}P_{2}-T_{1}P_{1}\right)\cdot R_{i}=0,\, i\in\{1,2\}$,
so that under the assumptions taken for $\hat{R}_{i}$ we obtain $\hat{R}_{1}=\hat{R}_{2}$.
These two equations imply
\begin{equation}
T_{2}\sqrt{\left(\hat{K}\cdot P_{2}\right)^{2}-P_{2}^{2}}-T_{1}\sqrt{\left(\hat{K}\cdot P_{1}\right)^{2}-P_{1}^{2}}=0.\label{eq:tdelayscalar2}
\end{equation}

Computing \eqref{eq:tdelayscalar1} - \eqref{eq:tdelayscalar2} we
obtain the expression for $\Delta S$ in a form that makes it easier
to examine: 
\begin{equation}
\Delta S=T_{2}\left[\left(\hat{K}\cdot P_{2}\right)-\sqrt{\left(\hat{K}\cdot P_{2}\right)^{2}-P_{2}^{2}}\right]-T_{1}\left[\left(\hat{K}\cdot P_{1}\right)-\sqrt{\left(\hat{K}\cdot P_{1}\right)^{2}-P_{1}^{2}}\right],\label{eq:tdelayfinal}
\end{equation}
since it is now evident that if the photons are null ($P_{i}^{2}=0$),
$\Delta S=0$, so there is no time delay. On the other hand, if $P_{i}^{2}\neq0$,
$\Delta S$ depends nontrivially in the photons' energies, and in
the experimental situation of a gamma ray burst, when usually one
of them has a much higher energy than the other, it is possible to
obtain a measurable time delay. To first order in the momenta, this
can only happen if there is non-metricity present, as shown by \eqref{eq:nonmetricnorms}.

\subsubsection{Specifying momentum constraints - 1st alternative}

We will now rework the problem with the following set of momentum
constraints:
\begin{equation}
\begin{array}{cccccc}
\mathcal{K}^{1}= & \ominus p^{1}\oplus(q^{1}\ominus k^{1}) & =0\qquad & \mathcal{K}^{2}= & \ominus p^{2}\oplus(k^{1}\ominus r^{1}) & =0\\
\mathcal{K}^{3}= & p^{1}\oplus(\ominus k^{2}\oplus q^{2}) & =0\qquad & \mathcal{K}^{4}= & p^{2}\oplus(\ominus r^{2}\oplus k^{2}) & =0
\end{array}
\end{equation}
Only the first two are changed through the diffeomorphisms $\mathcal{K}^{i}=L_{\ominus p^{i}}R_{p^{i}}\mathcal{K}_{\text{(orig)}}^{i}\equiv\lambda^{i}\left(\mathcal{K}_{\text{(orig)}}^{i}\right)$.
The relative locality operators corresponding to the altered constraints
are
\begin{equation}
\begin{array}{cccccccc}
W_{x_{1}}= & U_{0}^{p^{1}}V_{p^{1}}^{q^{1}} & \qquad & W_{u_{1}}= & -U_{0}^{p^{1}}U_{p^{1}}^{\ominus k^{1}}I^{k^{1}} & \qquad & W_{y_{1}}= & U_{0}^{p^{1}}\\
W_{x_{2}}= & U_{0}^{p^{2}}V_{p^{2}}^{k^{1}} & \qquad & W_{u_{2}}= & -U_{0}^{p^{2}}U_{p^{2}}^{\ominus r^{1}}I^{r^{1}} & \qquad & W_{y_{2}}= & U_{0}^{p^{2}},
\end{array}
\end{equation}
which all obey similar relations to the original ones, $W_{A_{i}}=U_{0}^{p^{i}}V_{p^{i}}^{0}W_{A_{i}}^{\text{(orig)}},\, A\in\left\{ x,\, u,\, y\right\} $.
We can then interpret the change in the relative locality relations
for these worldlines using gauge transformations, where the interaction
coordinates change as
\begin{equation}
z_{i}'=z_{i}\left(\frac{\partial\lambda^{i}}{\partial\mathcal{K}_{(\text{orig})}^{i}}\right)^{-1}=z_{i}d_{0}\left(L_{\ominus p^{i}}R_{p^{i}}\right)=z_{i}U_{0}^{p^{i}}V_{p^{i}}^{0},
\end{equation}
agreeing with the changes in $W_{A_{I}}$. When can we be sure that
gauge transformations such as this one do not affect the physics?
They tell us that after a change of vertex, one can obtain the same
relative locality relations given by the original vertex \textit{at
the cost of redefining the interaction coordinates $z_{(\alpha)}$}.
So a physical relation can only be affected if it depends on $z_{(\alpha)}$,
which is the case in our fundamental equation \eqref{eq:tdelay}.
Luckily, to 1st order in momenta for the original momentum constraints,
this dependence vanishes, so the physics of the problem should be
unaltered by any modification in the constraints that preserves this
cancellation.\\
\\
For our first attempt at a vertex change, 1st order calculations show
that the cancellation is indeed preserved - we see that $\left(W_{u_{1}}W_{x_{2}}^{-1}W_{y_{2}}W_{y_{4}}^{-1}\right)_{a}^{b}\approx\delta_{a}^{b}\approx\left(W_{y_{1}}W_{y_{3}}^{-1}W_{u_{3}}W_{x_{4}}^{-1}\right)_{a}^{b}$
so that the LHS of \eqref{eq:tdelay} still vanishes to 1st order,
and the expressions \eqref{eq:origmomenta} for the {}``uppercase
momenta'' remain valid (an equality can be readily verified to all
orders) - so \eqref{eq:tdelay} has exactly the same form.

\subsubsection{Specifying momentum constraints - 2nd alternative}

We will now consider this set of conservation laws:
\begin{equation}
\begin{array}{cccccc}
\mathcal{K}^{1}= & (\ominus k^{1}\oplus q^{1})\ominus p^{1} & =0\qquad & \mathcal{K}^{2}= & (\ominus r^{1}\oplus k^{1})\ominus p^{2} & =0\\
\mathcal{K}^{3}= & p^{1}\oplus(q^{2}\ominus k^{2}) & =0\qquad & \mathcal{K}^{4}= & p^{2}\oplus(k^{2}\ominus r^{2}) & =0
\end{array}
\end{equation}
All are modified in the same way with respect to the originals - the
order of addition within brackets is reversed. Computing the relative
locality operators:
\begin{equation}
\begin{array}{cccccccc}
W_{x_{1}}= & V_{0}^{p^{1}}U_{p^{1}}^{q^{1}} & \qquad & W_{u_{1}}= & -V_{0}^{p^{1}}V_{p^{1}}^{\ominus k^{1}}I^{k^{1}} & \qquad & W_{y_{1}}= & V_{0}^{p^{1}}\\
W_{x_{2}}= & V_{0}^{p^{2}}U_{p^{2}}^{k^{1}} & \qquad & W_{u_{2}}= & -V_{0}^{p^{2}}V_{p^{2}}^{\ominus r^{1}}I^{r^{1}} & \qquad & W_{y_{2}}= & V_{0}^{p^{2}}\\
W_{x_{3}}= & U_{0}^{\ominus p^{1}}V_{\ominus p^{1}}^{q^{2}} & \qquad & W_{u_{3}}= & -U_{0}^{\ominus p^{1}}U_{\ominus p^{1}}^{\ominus k^{2}}I^{k^{2}} & \qquad & W_{y_{3}}= & V_{0}^{p^{1}}\\
W_{x_{4}}= & U_{0}^{\ominus p^{2}}V_{\ominus p^{2}}^{k^{2}} & \qquad & W_{u_{4}}= & -U_{0}^{\ominus p^{2}}U_{\ominus p^{2}}^{\ominus r^{2}}I^{r^{2}} & \qquad & W_{y_{4}}= & V_{0}^{p^{2}}.
\end{array}
\end{equation}
The diffeomorphisms relating the new momentum constraints to the original
ones are
\begin{equation}
\begin{array}{ccccc}
\lambda^{1}= & L_{\ominus k^{1}\oplus q^{1}}L_{k^{1}\ominus q^{1}} & \qquad & \lambda^{2}= & L_{\ominus r^{1}\oplus k^{1}}L_{r^{1}\ominus k^{1}}\\
\lambda^{3}= & R_{q^{2}\ominus k^{2}}R_{\ominus q^{2}\oplus k^{2}} & \qquad & \lambda^{4}= & R_{k^{2}\ominus r^{2}}R_{\ominus k^{2}\oplus r^{2}},
\end{array}
\end{equation}
We cannot, however, apply the same gauge formalism as before, because
these diffeomorphisms do not respect $\lambda^{i}(0)=0$. They correspond
to legitimate changes in the conservation law imposed: $\mathcal{K}^{(1)}=0\Leftrightarrow p^{1}=\ominus k^{1}\oplus q^{1}$
but $\mathcal{K}^{(2)}=0\Leftrightarrow p^{1}=q^{1}\ominus k^{1}$,
which is obviously different if $\oplus$ is not commutative. The
fact that this is not a mere gauge transformation is further supported
by the fact that the $W_{A_{i}}$'s relations with the $W_{A_{i}}^{\text{(orig)}}$
are dependent on \textit{A}: for example $W_{x_{1}}=W_{x_{1}}^{\text{(orig)}}V_{q^{1}}^{p^{1}}U_{p^{1}}^{q^{1}}$,
$W_{u_{1}}=W_{u_{1}}^{\text{(orig)}}I^{\ominus k^{1}}U_{\ominus k^{1}}^{p^{1}}V_{p^{1}}^{\ominus k^{1}}$
and $W_{y_{1}}=W_{y_{1}}^{\text{(orig)}}$.\\
\\
Carrying on with the 1st order computation of \eqref{eq:tdelay},
though, we verify that $\left(W_{u_{1}}W_{x_{2}}^{-1}W_{y_{2}}W_{y_{4}}^{-1}\right)_{a}^{b}\approx\delta_{a}^{b}-T_{a}^{bc}p_{c}^{2}\approx\left(W_{y_{1}}W_{y_{3}}^{-1}W_{u_{3}}W_{x_{4}}^{-1}\right)_{a}^{b}$
so that the LHS vanishes once again, and the new uppercase momenta
are given by

\begin{eqnarray}
K_{1}^{b} & \approx & \hat{k}_{1}^{b}-m_{1}\hat{k}_{1}^{a}\Gamma_{a}^{cb}\hat{k}_{c}^{1}-E_{2}\hat{k}_{1}^{a}T_{a}^{bc}\hat{p}_{c}^{2}\approx\left[\left(\hat{k}_{1}U_{k^{1}}^{0}\right)U_{0}^{p^{2}}V_{p_{2}}^{0}\right]^{b}\nonumber \\
K_{2}^{b} & \approx & \hat{k}_{2}^{b}-m_{2}\hat{k}_{2}^{a}\Gamma_{a}^{bc}\hat{k}_{c}^{2}-E_{2}\hat{k}_{2}^{a}T_{a}^{bc}\hat{p}_{c}^{2}\approx\left[\left(\hat{k}_{2}V_{k^{2}}^{0}\right)U_{0}^{p^{2}}V_{p_{2}}^{0}\right]^{b}\nonumber \\
P_{1}^{b} & \approx & \hat{p}_{1}^{b}-E_{1}\hat{p}_{1}^{a}\Gamma_{a}^{bc}\hat{p}_{c}^{1}-E_{2}\hat{p}_{1}^{a}T_{a}^{bc}\hat{p}_{c}^{2}\approx\left[\left(\hat{p}_{1}V_{p^{1}}^{0}\right)U_{0}^{p^{2}}V_{p_{2}}^{0}\right]^{b}\nonumber \\
P_{2}^{b} & \approx & \hat{p}_{2}^{b}-E_{2}\hat{p}_{2}^{a}\Gamma_{a}^{cb}\hat{p}_{c}^{2}-E_{2}\hat{p}_{2}^{a}T_{a}^{bc}\hat{p}_{c}^{2}\approx\left[\left(\hat{p}_{2}U_{p^{2}}^{0}\right)U_{0}^{p^{2}}V_{p_{2}}^{0}\right]^{b}.
\end{eqnarray}
We can multiply \eqref{eq:tdelay} by $\left(U_{0}^{p^{2}}V_{p_{2}}^{0}\right)^{-1}$
to recover the result that, to first order, the uppercase momenta
depend only on the lowercase ones. They are not the same as \eqref{eq:origmomenta}:
the difference in each one is a term proportional to $T_{a}^{bc}\mathbf{p}_{c}^{i}$
where $\mathbf{p}\in\{k,\, p\}$. But, since \cite{key-2}'s calculation
of the time delay does not include torsion, we would still obtain
the same result as the paper does.

\subsubsection{Specifying momentum constraints - 3rd alternative}

We will now consider a similar modification of momentum constraints
as in 3.3.3, but only commuting the momenta within brackets in the
first two:
\begin{equation}
\begin{array}{cccccc}
\mathcal{K}^{1}= & (q^{1}\ominus k^{1})\ominus p^{1} & =0\qquad & \mathcal{K}^{2}= & (k^{1}\ominus r^{1})\ominus p^{2} & =0\\
\mathcal{K}^{3}= & p^{1}\oplus(q^{2}\ominus k^{2}) & =0\qquad & \mathcal{K}^{4}= & p^{2}\oplus(k^{2}\ominus r^{2}) & =0
\end{array}
\end{equation}
All the corresponding relative locality operators have been computed
in the previous sections, so we will just state the most remarkable
result of computing the time delay equation: we obtain $\left(W_{u_{1}}W_{x_{2}}^{-1}W_{y_{2}}W_{y_{4}}^{-1}\right)_{a}^{b}\approx\delta_{a}^{b}+T_{a}^{bc}p_{c}^{1}$
but $\left(W_{y_{1}}W_{y_{3}}^{-1}W_{u_{3}}W_{x_{4}}^{-1}\right)_{a}^{b}\approx\delta_{a}^{b}-T_{a}^{bc}p_{c}^{2}$,
so the LHS of \eqref{eq:tdelay} does not vanish to 1st order in the
momenta. Instead we get a term $z_{1}T_{a}^{bc}\left(p^{1}+p^{2}\right)_{c}$,
which depends on an interaction coordinate. This is problematic, because
it means the calculation we followed in 3.3.1 to obtain the delay
would not (in the full case with torsion) allow us to obtain a result
in terms of diffeomorphism-covariant quantities only; there appears
to be a legitimate change in the physics following the change of vertex
in this case.\\
\\
However, since the offending term is also proportional to the torsion,
in the torsion-free momentum space considered for the original time
delay calculation we can still carry on with the derivation in a covariant
way. Computing the parallel transported momenta:is 
\begin{eqnarray}
K_{1}^{b} & \approx & \hat{k}_{1}^{b}-m_{1}\hat{k}_{1}^{a}\Gamma_{a}^{bc}\hat{k}_{c}^{1}\approx\left(\hat{k}_{1}V_{k^{1}}^{0}\right)^{b}\nonumber \\
K_{2}^{b} & \approx & \hat{k}_{2}^{b}-m_{2}\hat{k}_{2}^{a}\Gamma_{a}^{bc}\hat{k}_{c}^{2}-E_{2}\hat{k}_{2}^{a}T_{a}^{bc}\hat{p}_{c}^{2}\approx\left(\hat{k}_{2}V_{k^{2}}^{0}U_{0}^{p^{2}}V_{p_{2}}^{0}\right)^{b}\nonumber \\
P_{1}^{b} & \approx & \hat{p}_{1}^{b}-E_{1}\hat{p}_{1}^{a}\Gamma_{a}^{bc}\hat{p}_{c}^{1}-E_{2}\hat{p}_{1}^{a}T_{a}^{bc}\hat{p}_{c}^{2}\approx\left(\hat{p}_{1}V_{p^{1}}^{0}U_{0}^{p^{2}}V_{p_{2}}^{0}\right)^{b}\nonumber \\
P_{2}^{b} & \approx & \hat{p}_{2}^{b}-E_{2}\hat{p}_{2}^{a}\Gamma_{a}^{bc}\hat{p}_{c}^{2}\approx\left(\hat{p}_{2}V_{p^{2}}^{0}\right)^{b}
\end{eqnarray}
Again, the difference between these and the ones resulting from the
original choice of momentum constraints is proportional to the torsion.
Thus we find that to first order, if momentum space is torsion-free,
none of the vertex changes considered alter the physics. Indeed, this
statement can be generalized to any vertex change - to this order,
momentum addition is associative, and without torsion it is also commutative,
which means all forms of writing $\mathcal{K}_{b}=\overset{N}{\underset{J=1}{\oplus}}k_{b}^{J}=0$
become equivalent.

\begin{onehalfspace}

\section{Multiparticle systems coupled to 2+1 gravity}
\end{onehalfspace}

In this section we will outline the known picture of the phase space
of a model of $N$ point particles coupled to 2+1 Einstein gravity,
which is described in detail in \cite{key-8}. In the following, $\mathcal{M}$
is the spacetime manifold, and index notation is as follows: $a,\, b,\, c...$
are tangent space indices; $i,\, j...$ are Lie algebra indices and
$\pi_{1},\,\pi_{2}...$ are particle labels.

\subsection{Phase space of multiple point particle systems}

The starting point to understanding the construction of the multiparticle
phase space is the result shown in \cite{key-10} for the spacetime
metric induced by a point particle of mass \textit{m} and spin \textit{s}:
\begin{equation}
ds^{2}=-(dt+4Gs\, d\phi)^{2}+dr^{2}+r^{2}\left[(1-4Gm)d\phi\right]^{2}\label{eq:spctgeometry}
\end{equation}
Physically, the geometry represented by this metric is flat (as can
be seen from computing the Riemann tensor), but with a \textit{time
offset} of $8\pi Gs$ and a \textit{conical deficit angle} of $8\pi Gm$,
which means that, even though there is no local gravitational force
acting on the particles, they still interact with each other due to
the topological effects, a peculiar characteristic of this model.
This geometry also imposes a maximum bound on the total mass of the
system: the sum of all deficit angles cannot exceed $2\pi$, so we
have that $M=\sum_{\pi=1}^{N}m_{\pi}<\frac{1}{4G}$.

\begin{figure}[H]
\noindent \begin{centering}
\includegraphics{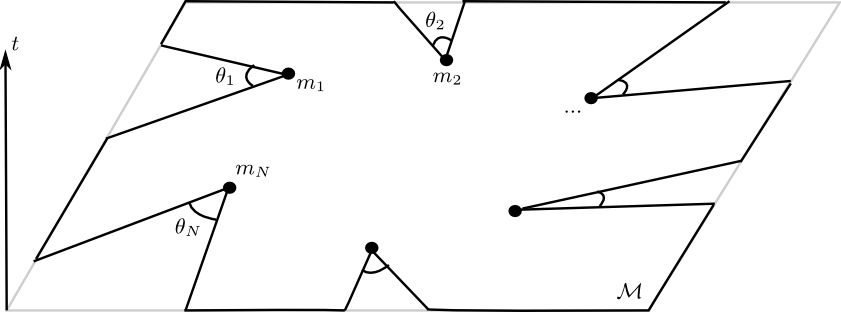}
\par\end{centering}

\caption{Geometry of the 2+1 multiparticle system coupled to gravity.}
\end{figure}

We use the first-order formalism of Einstein gravity (see Appendix
for a brief review; \cite{key-9} for a more detailed one) with the
frame fields and the spin connection as dynamical variables, which
suggests an infinite dimensional phase space. However, this space
can be reduced via application of constraints and quotienting out
of gauge symmetries, to an extent where it is actually finite-dimensional,
with a set of dynamical variables that are essentially the positions
and momenta of the point particles. Throughout, we will work in the
ADM formalism \cite{key-11} for gravity, by splitting $\mathcal{M}\approx\mathbb{R}\times\Sigma$,
where $\Sigma$ has the topology of $\mathbb{R}^{2}$ with $N$ {}``holes''
cut out, corresponding to the particles.

\subsubsection*{The need to redefine phase space variables}

To better understand the formalism at hand we will start by considering
the problem without gravity. We then have a system of free particles,
which can be parameterized by the variables $x_{\pi}=x_{\pi}^{i}\gamma_{i}$
and $p_{\pi}=p_{\pi}^{i}\gamma_{i}$ (using the isomorphism $\mathbb{M}^{2,1}\approx sl(2)$)
satisfying the Poisson brackets $\left\{ p_{\pi}^{i},x_{\pi}^{j}\right\} =\eta^{ij}$.
The phase space is $6N$-dimensional, and the \textit{physical} phase
space is the submanifold defined by the mass shell and positive energy
conditions:
\begin{equation}
\frac{1}{2}Tr\left(p_{\pi}^{2}\right)=-m_{\pi}^{2},\qquad p_{\pi}^{0}=\frac{1}{2}Tr\left(p_{\pi}^{2}\right)>0.
\end{equation}
The evolution equations are similar to those derived in Section 3.1,
but from the traditional spacetime perspective,
\begin{equation}
\dot{p}_{\pi}=0,\qquad\dot{x}_{\pi}=\mathcal{N}_{\pi}p_{\pi}.
\end{equation}
Dots represent derivatives with respect to ADM time, which can of
course be freely reparameterized. This is the gauge freedom of the
system, which means we can fix $\mathcal{N}_{\pi}$ to be whatever
we want. However, there is a gauge restriction to be imposed: that
at a given ADM time $t$, all particles are located in the same spacelike
surface $\Sigma(t)$, which translates to 
\begin{equation}
x_{\pi_{2}}-x_{\pi_{1}}\text{ is spacelike }\forall\pi_{1},\pi_{2}\in\{1,...,N\}.\label{eq:spacelikeconstraint}
\end{equation}
We are still left with the Poincaré symmetry, which can be partially
gauged away by moving to the center of mass frame (suppose such a
frame exists by excluding the case where all the particles are massless),
characterized by the expressions for total momentum and total angular
momentum of the system
\begin{equation}
P=\sum_{\pi}p_{\pi}=M\gamma_{0},\qquad J=\frac{1}{2}\sum_{\pi}\left[p_{\pi},\, x_{\pi}\right]=S\gamma_{0},\label{eq:CMframe}
\end{equation}
leaving only time translations and spatial rotations as ungauged symmetries,
and a phase space of dimension $6N-4$. However, this raises a problem.
The constraints \eqref{eq:CMframe} are second class, which is undesirable
for a Hamiltonian description. One way of solving the problem is to
consider a different picture of phase space for which \eqref{eq:CMframe}
become 1st class constraints. This is done by reparameterizing their
solutions. The convenient reparameterization is the triangulation\textit{
}picture, which we will describe first for the model without gravity
and then adapt to the gravity model.

\subsubsection*{Triangulations}

Consider the foliation of spacetime given by \eqref{eq:spacelikeconstraint},
$\mathcal{M}=\mathbb{R}\times\Sigma(t)$. A \textit{triangulation}
$\Gamma$ is defined by a set of oriented links between particles,
$\lambda$, and links from particles to infinity, $\eta$, which split
the ADM surface into finite polygons $\Delta_{\lambda}$ and polygons
with a vertex at infinity, $\Delta_{\eta}$. The preferred orientation
for the links is chosen \textit{a priori} and all particles are connected
by links in both directions, so that the triangulation is split in
the sets of positive-oriented, negative-oriented and infinity-bound
(external) links, $\Gamma=\Gamma_{+}\cup\Gamma_{-}\cup\Gamma_{\infty}=\Gamma_{0}\cup\Gamma_{\infty}$.
Also define the set of links that \textit{end} at a particle $\pi$,
$\Gamma_{\pi}$, and the set of links that \textit{begin} at $\pi$,
$\Gamma_{-\pi}$. The particle in which $\lambda$ ends is called
$\pi_{\lambda}$, and the particle in which $\lambda$ begins is called
$\pi_{-\lambda}$.

\begin{figure}[H]
\noindent \begin{centering}
\includegraphics{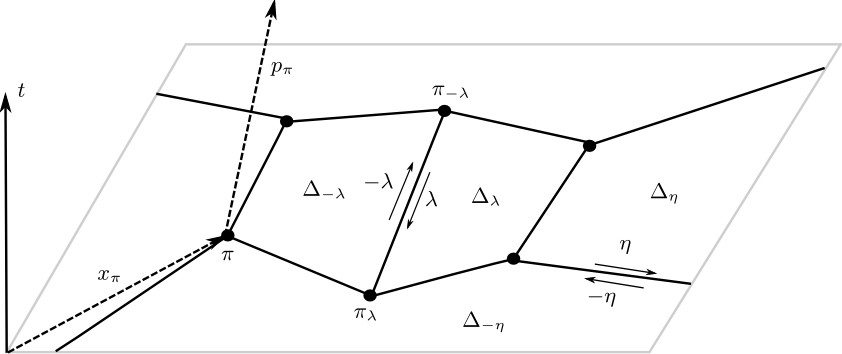}
\par\end{centering}

\caption{Triangulation of a Minkowski ADM surface.}
\end{figure}

We define the \textit{relative momenta} $q_{\lambda}$, related to
the particles' momenta by the equations
\begin{equation}
p_{\pi}=\sum_{\lambda\in\Gamma_{\pi}}q_{\lambda},\qquad q_{-\lambda}=-q_{\lambda},\quad\lambda\in\Gamma_{0},
\end{equation}
the \textit{relative positions} $z_{\lambda}$ given by
\begin{equation}
z_{\lambda}=x_{\pi_{\lambda}}-x_{\pi_{-\lambda}},
\end{equation}
and the variables associated with external links%
\footnote{$\gamma(\phi)=\cos(\phi)\gamma_{1}+\sin(\phi)\gamma_{2}$%
},
\begin{equation}
q_{\eta}=-M_{\eta}\gamma_{0},\quad z_{\eta}=\gamma\left(\phi_{\eta}\right),\quad T_{\eta}=x_{\pi_{-\lambda}}^{0},\quad\eta\in\Gamma_{\infty},
\end{equation}
where $\phi_{\eta}$ is the polar angle of $\eta$ and the $T_{\eta}$
represent the time coordinates of particles at the ends of the external
links as measured by a clock at infinity.\\
\\
With this setup, the CM frame constraints \eqref{eq:CMframe} are
automatically solved if
\begin{equation}
\sum_{\eta\in\Gamma_{\infty}}M_{\eta}=M,\qquad\sum_{\lambda\in\Gamma_{+}}L_{\lambda}=\sum_{\lambda\in\Gamma_{+}}\frac{1}{4}Tr\left(\left[q_{\lambda},\, z_{\lambda}\right]\gamma_{0}\right)=S.\label{eq:nonGCMconst}
\end{equation}
\eqref{eq:nonGCMconst} are first class - the primary goal of the
triangulation procedure. It is important to notice the link variable
definitions have in them several redundancies, which can be accounted
for as extra kinematical constraints. The end result is that the particles'
momenta in the center of mass frame can be completely specified in
terms of link variables, and while absolute positions can only be
derived up to an overall translation from the relative ones, the angular
momentum constraint fixes them up to a time translation $x_{\pi}\rightarrow x_{\pi}+\tau\gamma_{0}$
and rotations of the external links $\phi_{\eta}\rightarrow\phi_{\eta}+\epsilon_{\eta}$
that can be achieved by smooth deformation of the ADM surface.\\
\\
Introducing the kinematical and dynamical constraints into the ordinary
free-particle Hamiltonian, the resulting nonzero Poisson brackets
of the system are
\begin{equation}
\begin{array}{cccccccc}
\left\{ T_{\eta,}\, M_{\eta}\right\}  & = & 1 & \qquad & \left\{ L_{\eta,}\,\phi_{\eta}\right\}  & = & 1\\
\left\{ q_{\lambda,}^{i}\, z_{\lambda}^{j}\right\}  & = & \eta^{ij} & \qquad & \left\{ q_{-\lambda,}^{i}\, z_{\lambda}^{j}\right\}  & = & -\eta^{ij} & =\left\{ q_{\lambda,}^{i}\, z_{-\lambda}^{j}\right\} .
\end{array}
\end{equation}

\subsubsection*{Phase space in the presence of gravity}

The biggest geometrical difference when applying the above formalism
to gravity is the presence of the conical deficits. This means spacetime
is not simply connected, so it is no longer possible to describe the
system by global Minkowski coordinates - instead, the ADM surface
is covered by an atlas whose charts are Minkowski, since geometry
is still flat everywhere outside the particles.

\begin{figure}[H]
\noindent \begin{centering}
\includegraphics{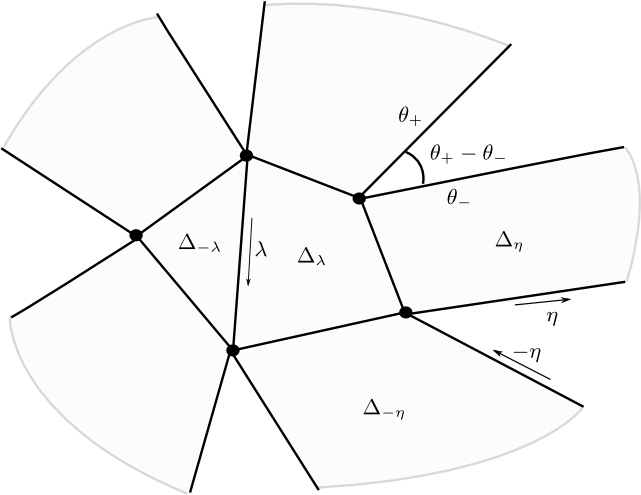}
\par\end{centering}

\caption{Triangulation of an ADM surface with gravity inducing conical defects.}

\end{figure}

The Minkowski charts $\Phi_{\Delta}$ are chosen so that each one
covers a polygon $\Delta$, and for the atlas to be well-defined there
must be diffeomorphisms relating two different sets of coordinates
in intersecting patches, the transfer functions. Since the intersection
of two polygons is a link, the transfer functions are labeled by the
links, and necessarily correspond to isometries of Minkowski space
mapping $-\lambda$ to $\lambda$, of which we will only consider
the Lorentz rotations $g_{\lambda}\in SL(2)$. They relate space coordinates
in the two orientations of a link:
\begin{equation}
z_{-\lambda}=-g_{\lambda}z_{\lambda}g_{\lambda}^{-1}.
\end{equation}
Since the relative positions and the transfer functions contain all
information about the shape of the polygons and how to {}``glue''
them together to form the ADM surface, they effectively encode spacetime
geometry - up to redundancy transformations that correspond to diffeomorphism
invariance.\\
\\
All link variables given for flat space are still definable for the
current case, except for the relative momenta, which are now related
to the transfer functions. The relation is given by considering a
small-$G$ approximation of the problem and checking consistency of
the definition with the flat-space formulas given above: one can write
(using the exponential map $\exp:\: sl(2)\rightarrow SL(2)$)%
\footnote{Factor of $4\pi G$ for dimensional correctness.%
}
\begin{equation}
g_{\lambda}=e^{4\pi Gq_{\lambda}}\approx1+4\pi Gq_{\lambda}+\mathcal{O}(q_{\lambda}^{2}),\quad4\pi G\cdot\left|q_{\lambda}\right|\ll1.
\end{equation}
Additionally, the transfer functions in external links are defined
to allow satisfaction of the center of mass constraints, as before:
\begin{equation}
g_{\pm\eta}=e^{\mp4\pi GM_{\eta}\gamma_{0}},\;\eta\in\Gamma_{\infty}.
\end{equation}
More remarkable is the assertion that the \textit{absolute }momenta
of the point particles can be deduced from the transfer functions,
through structures called \textit{holonomies}. The holonomy is defined
by considering the parallel transport of a spacetime vector along
a closed path around a particle. With the present geometry, parallel
transport is trivial within polygons, but picks up a Lorentz rotation
from a transfer function at each conical deficit, which adds up to
a rotation of $8\pi Gm$ along a timelike axis. From this Lorentz
group transformation one can derive the 3-momentum of the particle.
Put more formally, the holonomy of a particle $\pi$ evaluated at
a polygon $\Delta$ is the product of the transfer functions,
\begin{equation}
u_{\pi,\Delta}=\prod_{\lambda\in\Gamma_{\pi,\Delta}}g_{\lambda}.
\end{equation}
Under Lorentz rotations of the charts parametrized by $\left\{ h_{\Delta}\right\} _{\Delta\in\Gamma}$,
where $h_{\Delta}\in SL(2)$, the transfer functions transform as
\begin{equation}
g_{\lambda}\rightarrow h_{\Delta_{-\lambda}}g_{\lambda}h_{\Delta_{-\lambda}}^{-1},
\end{equation}
and hence, the holonomy transforms correctly under \textit{SL}(2),
thus holonomies at different polygons only differ by the corresponding
Lorentz rotations relating the corresponding charts,
\begin{equation}
u_{\pi,\Delta}\rightarrow h_{\Delta}u_{\pi,\Delta}h_{\Delta}^{-1},\:\forall\Delta\Rightarrow u_{\pi,\Delta_{\lambda}}=g_{\lambda}^{-1}u_{\pi,\Delta_{-\lambda}}g_{\lambda},\:\forall\lambda.
\end{equation}
3-momentum is deduced from the canonical projection $SL(2)\rightarrow sl(2)$
and expressed as follows: 
\begin{equation}
u_{\pi}=\mathcal{U}_{\pi}\mathbf{1}+4\pi Gp_{\pi}^{i}\gamma_{i},\label{eq:constitutive}
\end{equation}
where $\mathcal{U}_{\pi}$ is the \textit{holonomy scalar.} The factor
$4\pi G$ is not only necessary for dimensional correctness, it fixes
the correct mass shell constraint for the model - which we derive
by demanding the holonomy to be a rotation of $8\pi Gm$ along the
timelike axis as above. The dynamical constraints are then 
\begin{equation}
\mathcal{U}_{\pi}=\cos\left(4\pi Gm_{\pi}\right),\qquad p_{\pi}^{0}=\frac{1}{2}Tr\left(u_{\pi}\gamma^{0}\right)>0.
\end{equation}
The fact that $u_{\pi}\in SL(2)\Rightarrow\det\left(u_{\pi}\right)=1$
gives an extra relation $(4\pi G)^{2}\frac{1}{2}Tr\left(\left(p_{\pi}\right)^{2}\right)-\mathcal{U}_{\pi}^{2}=-1$,
which allows us to rewrite the mass shell constraint as 
\begin{eqnarray}
\frac{1}{2}Tr\left(\left(p_{\pi}\right)^{2}\right) & = & \eta_{ij}p_{\pi}^{i}p_{\pi}^{j}=-\frac{\sin^{2}\left(4\pi Gm_{\pi}\right)}{\left(4\pi G\right)^{2}},\label{eq:masshellconst}
\end{eqnarray}
with the correct pre-factor of $4\pi G$ giving the usual flat space
mass shell relation when $G\rightarrow0$.\\
\\
The introduction of deficit angles $\theta_{\eta}^{\pm}$ and time
offsets $\tau_{\Delta}\left(\phi_{\eta}\right)$ of the new geometry
leads to updated definitions for the external link variables,
\begin{equation}
\begin{array}{ccccccccccc}
M_{\eta} & = & \frac{1}{8\pi G}\left(\theta_{\eta}^{+}-\theta_{\eta}^{-}\right), & \, & z_{\pm\eta} & = & \pm\gamma\left(\phi_{\eta}+\theta_{\eta}^{\pm}\right), & \, & T_{\eta} & = & x_{\pi_{-\lambda}}^{0}-\tau_{\Delta}\left(\phi_{\eta}\right),\end{array}
\end{equation}
new CM frame constraints which remain first class,
\begin{equation}
\sum_{\eta\in\Gamma_{\infty}}M_{\eta}=M,\qquad\sum_{\lambda\in\Gamma_{+}}L_{\lambda}=\sum_{\lambda\in\Gamma_{+}}\frac{1}{16\pi G}Tr\left(\left(g_{\lambda}z_{\lambda}g_{\lambda}^{-1}-z_{\lambda}\right)\gamma^{0}\right)=S,
\end{equation}
and new kinematical constraints with terms proportional to the masses
and spins of the particles. More importantly, the new nonvanishing
Poisson brackets are 
\begin{equation}
\begin{array}{cccccccc}
\left\{ T_{\eta,}\, M_{\eta}\right\}  & = & 1 & \qquad & \left\{ L_{\eta,}\,\phi_{\eta}\right\}  & = & 1\\
\left\{ g_{\lambda,}\, z_{\lambda}^{i}\right\}  & = & 4\pi Gg_{\lambda}\gamma^{i} & \qquad & \left\{ g_{-\lambda,}\, z_{\lambda}^{i}\right\}  & = & -4\pi G\gamma^{i}g_{\lambda} & =\left\{ g_{\lambda,}\, z_{-\lambda}^{i}\right\} \\
\left\{ z_{\lambda,}^{i}\, z_{\lambda}^{j}\right\}  & = & 8\pi G\epsilon_{\;\,\,\, k}^{ij}z_{\lambda}^{k}
\end{array}\label{eq:poissonbracketsG}
\end{equation}
with the key difference of having a noncommutative Poisson algebra
between relative positions.\\
\\
In \cite{key-8}, it is described in detail how to derive the above
outlined phase space picture from Einstein gravity, which consists
essentially in writing the first-order action \eqref{eq:firstorderaction}
in ADM form and adding extra particle terms and kinematic constraints
to accurately reproduce the algebra \eqref{eq:poissonbracketsG}.
We just note that it is straightforward to find the frame field and
spin connection one forms that describe the metric \eqref{eq:spctgeometry},
\begin{eqnarray}
e_{a} & = & \left(\partial_{a}t+4Gs\partial_{a}\phi\right)\gamma_{0}+\partial_{a}r\gamma(\phi)+\left(1-4Gm\right)r\partial_{a}\phi\gamma'(\phi),\nonumber \\
\omega_{a} & = & -2GM\partial_{a}\phi\gamma_{0},
\end{eqnarray}
and the holonomy scalar is computed in terms of the spin connection
in the boundary of the {}``particle hole'' $B_{\pi}$:
\begin{equation}
\mathcal{U}_{\pi}=\frac{1}{2}Tr\left(\mathcal{P}exp\int_{B_{\pi}}d\phi\wedge\omega\right).
\end{equation}

\subsection{The relative locality interpretation}

We will now draw the picture of phase space in the 2+1 multiparticle
system from a different perspective: that of relative locality. In
order to do so, we need to characterize the geometry of momentum space.
Two pieces of data are needed:
\begin{itemize}
\item the metric of momentum space;
\item the combination rule between two momenta.
\end{itemize}
The metric is easily obtainable from the description of momentum as
derived from a particle's holonomy: the constitutive relation \eqref{eq:constitutive}
can be rewritten in matricial form (going back to the notation of
lower indices for momenta and dropping the $\pi,\,\Delta$ subscripts)
\begin{equation}
u=4\pi G\left[\begin{matrix}\frac{\mathcal{U}}{4\pi G}-p_{2} & p_{1}+p_{0}\\
p_{1}-p_{0} & \frac{\mathcal{U}}{4\pi G}+p_{2}
\end{matrix}\right].
\end{equation}
The condition that $u\in SL(2)$ can be rewritten in the new notation
as
\begin{eqnarray*}
(4\pi G)^{2}\eta^{ab}p_{a}p_{b}-\mathcal{U}^{2} & = & -1.
\end{eqnarray*}
Defining the coordinate $p_{3}=\frac{\mathcal{U}}{4\pi G}$, and $\eta^{AB}=\text{diag}(-1,+1,+1,-1)$,
we see that this equation is simply the embedding equation of 3-dimensional
anti-de Sitter space in Minkowski space $\mathbb{M}^{2,2}$, $\eta^{AB}p_{A}p_{B}=-\rho^{2}$,
where the curvature radius is
\begin{equation}
\rho=\frac{1}{4\pi G},
\end{equation}
and the metric of momentum space is readily written in the quasi-cartesian
coordinate system we are working in:
\begin{equation}
g^{ab}(p)=\eta^{ab}-\frac{\eta^{ar}\eta^{bs}p_{r}p_{s}}{\eta^{rs}p_{r}p_{s}+\rho^{2}}.
\end{equation}
To derive a momentum addition rule we just have to note that putting
together two holonomies $u_{\pi_{1},\Delta_{1}},\, u_{\pi_{2},\Delta_{2}}$
(which corresponds physically to the absorption of a particle by the
other), the resultant holonomy is simply the group product of the
two: $u_{\pi_{1}\oplus\pi_{2},\Delta_{1}\oplus\Delta_{2}}=u_{\pi_{1},\Delta_{1}}\cdot u_{\pi_{2},\Delta_{2}}$.
If the corresponding momenta are $p,q$ respectively, we obtain (denoting
$p^{2}=\eta^{ab}p_{a}p_{b}$ and $p\cdot q=\eta^{ab}p_{a}q_{b}$)
\begin{equation}
u_{\pi_{1}\oplus\pi_{2},\Delta_{1}\oplus\Delta_{2}}=4\pi G\left[\begin{matrix}\frac{\mathcal{U}_{\pi_{1}\oplus\pi_{2}}}{4\pi G}-\left(p\oplus q\right)_{2} & \left(p\oplus q\right)_{1}+\left(p\oplus q\right)_{0}\\
\left(p\oplus q\right)_{1}-\left(p\oplus q\right)_{0} & \frac{\mathcal{U}_{\pi_{1}\oplus\pi_{2}}}{4\pi G}+\left(p\oplus q\right)_{2}
\end{matrix}\right],
\end{equation}
where\textit{
\begin{eqnarray}
\mathcal{U}_{\pi_{1}\oplus\pi_{2}}=\sqrt{1+(4\pi G)^{2}(p\oplus q)^{2}} & = & \sqrt{1+(4\pi G)^{2}p^{2}}\sqrt{1+(4\pi G)^{2}q^{2}}-(4\pi G)^{2}p\cdot q\\
(p\oplus q)_{a} & = & \sqrt{1+(4\pi G)^{2}p^{2}}\, q_{a}+\sqrt{1+(4\pi G)^{2}q^{2}}\, p_{a}+4\pi G\eta_{ad}\epsilon^{dbc}p_{b}q_{c}.
\end{eqnarray}
}An important observation is that, since the combination rule is effectively
group multiplication in $SL(2)$, it is \textit{associative} - the
Riemann tensor \eqref{eq:riemann} vanishes. The connection is now
obtained from the definition \eqref{eq:connection}, with the result
\begin{eqnarray}
\Gamma_{c}^{ab}(p) & = & -(4\pi G)^{2}p_{c}g^{ab}-4\pi G\epsilon^{abd}\frac{(4\pi G)^{2}p_{d}p_{c}+\eta_{dc}}{\sqrt{1+(4\pi G)^{2}\eta^{rs}p_{r}p_{s}}}\nonumber \\
 & = & \left\{ _{\,\, c}^{ab}\right\} +\frac{1}{2}T_{c}^{ab},
\end{eqnarray}
where
\begin{equation}
\left\{ _{\,\, c}^{ab}\right\} =-(4\pi G)^{2}p_{c}g^{ab}
\end{equation}
are the Christoffel symbols of the $AdS_{3}$ metric and 
\begin{equation}
T_{c}^{ab}=-8\pi G\epsilon^{abd}\frac{(4\pi G)^{2}p_{d}p_{c}+\eta_{dc}}{\sqrt{1+(4\pi G)^{2}\eta^{rs}p_{r}p_{s}}}\label{eq:2+1torsion}
\end{equation}
is the torsion%
\footnote{Also notice that, fortuitously, $T^{iab}+T^{iba}=0$, so the corresponding
term in the decomposition \eqref{eq:decomp} vanishes.%
}. We find that the non-metricity tensor vanishes, hence the connection
is compatible with the metric.\\
\\
Lastly, we can re-obtain the mass shell constraint \eqref{eq:masshellconst}
using its definition in the relative locality context, \eqref{eq:relocshell}.
The geodesic distance $D(p,0)$ is computed in the center of mass
frame, by attempting a solution of the form $p_{a}(\tau)=\left(p_{0}(\tau),\,0,\,0\right),\; p_{0}(0)=0,\, p_{0}(1)=p$
to the geodesic equations,
\begin{equation}
\ddot{p}_{a}(\tau)+\Gamma_{a}^{bc}(p(\tau))\dot{p}_{b}(\tau)\dot{p}_{c}(\tau)=0.
\end{equation}
The \textit{ansatz} reduces them to one single ODE:
\begin{equation}
\ddot{p}_{0}+\frac{p_{0}\dot{p}_{0}^{2}}{\rho^{2}-p_{0}^{2}}=0.\label{eq:ode}
\end{equation}
The solution of \eqref{eq:ode} respecting the boundary conditions
which tends to the flat space linear form when $\frac{p}{\rho}\rightarrow0$
is
\begin{equation}
p_{0}(\tau)=\rho\sin\left(\tau\left(\frac{\pi}{2}-\arctan\sqrt{\left(\frac{\rho}{p}\right)^{2}-1}\right)\right)
\end{equation}
 and the geodesic length is obtained by computing the integral
\begin{equation}
m=\mathcal{L}=\int_{0}^{1}d\tau\sqrt{-g^{ab}(p(\tau))\dot{p}_{a}(\tau)\dot{p}_{b}(\tau)}=\int_{0}^{1}d\tau\sqrt{\frac{\dot{p}_{0}^{2}}{1-\frac{p_{0}^{2}}{\rho^{2}}}}=\frac{\rho}{2}\left(\pi-2\arctan\sqrt{\left(\frac{\rho}{p}\right)^{2}-1}\right),
\end{equation}
which does satisfy $\rho^{2}\sin^{2}\left(\frac{m}{\rho}\right)=p_{0}^{2}=-\eta^{ab}p_{a}p_{b}$,
\eqref{eq:masshellconst}.\\
\\
The phase space of the 2+1 multiparticle system coupled to gravity
is then described as follows: the dynamical variables are the particle's
holonomy momenta $p_{\pi}\in\mathcal{P}\approx AdS_{3}$, where the
curvature radius of $\mathcal{P}$ is $\rho=\frac{1}{4\pi G}$, and
their dual spacetime coordinates $x_{\pi}\in T_{p_{\pi}}^{*}\mathcal{P}\equiv\mathcal{X}\left(p_{\pi}\right)$
with the geometry of spacetime given by \eqref{eq:spctgeometry}.
The existence of torsion in momentum space results from momentum addition
being essentially the $SL(2)$ group rule, and the fact that it is
curved explains the noncommutative structure of spacetime made evident
in the algebra \eqref{eq:poissonbracketsG}. The constraints on phase
space are the mass shell constraints \eqref{eq:masshellconst} and
the positive energy constraint $p_{0}>0$, along with the 2nd class
kinematical constraints to the CM frame, which can be made 1st class
by the triangulation procedure.

\subsection{Gamma ray bursts in 2+1 dimensions}

We will now carry out the derivation of the time delay between photon
emission and reception for the 2+1 multiparticle system. The calculation
follows the same steps performed in Section 3.3, and there are two
observations that make our life easier:
\begin{itemize}
\item Even though we are now working in a different physical regime than
in Section 3.3 ($\hbar$ is still neglected but $G$ is present),
the fact that spacetime geometry is everywhere flat apart from the
conical deficits at the particles' locations means the kinematic relations
\eqref{eq:kin_id} are still valid, as long as we redefine $\hat{k}_{i}=\frac{k_{i}}{\left|k_{i}\right|}$
according to the mass shell constraints \eqref{eq:masshellconst}.
\item The fact that $\oplus$ is associative results in useful formulas
for the parallel transport operators:
\begin{equation}
U_{q}^{p}=U_{q}^{0}U_{0}^{p};\qquad V_{q}^{p}=V_{q}^{0}V_{0}^{p}.\label{eq:associativeformulas}
\end{equation}
Proof is as follows: $U_{q}^{0}U_{0}^{p}=U_{q\oplus0}^{0}U_{\ominus p\oplus p}^{p}=d_{0}L_{q}\cdot d_{p}L_{\ominus p}=d_{p}\left(L_{q}L_{\ominus p}\right)=d_{p}L_{q\ominus p}=U_{q}^{p}$,
where the associativity of $\oplus$ is used to assert that $L_{q}L_{\ominus p}=L_{q\ominus p}$.
Calculation for $V$ is entirely analogous.
\end{itemize}
We can also compute the parallel transport operators \textit{U}, \textit{V},
\textit{I} according to the present combination rule:
\begin{eqnarray}
\left(U_{p\oplus q}^{q}\right)_{a}^{b} & = & \sqrt{1+(4\pi G)^{2}p^{2}}\,\delta_{a}^{b}-4\pi G\eta_{ad}\epsilon^{dbc}p_{c}+(4\pi G)^{2}\frac{p_{a}\eta^{bc}q_{c}}{\sqrt{1+(4\pi G)^{2}q^{2}}}\nonumber \\
\left(V_{p\oplus q}^{p}\right)_{a}^{b} & = & \sqrt{1+(4\pi G)^{2}q^{2}}\,\delta_{a}^{b}+4\pi G\eta_{ad}\epsilon^{dbc}q_{c}+(4\pi G)^{2}\frac{q_{a}\eta^{bc}p_{c}}{\sqrt{1+(4\pi G)^{2}p^{2}}}\nonumber \\
\left(I^{p}\right)_{a}^{b} & = & -\delta_{a}^{b}.
\end{eqnarray}
We will now compute the time delay $\Delta S$ using the momentum
constraints as they are given in \cite{key-2}. The formulas \eqref{eq:associativeformulas}
greatly simplify the relative locality operators:
\begin{equation}
\begin{array}{cccccccc}
W_{x_{1}}= & V_{0}^{q^{1}} & \qquad & W_{u_{1}}= & V_{0}^{p^{1}}U_{p^{1}}^{0}V_{0}^{k^{1}} & \qquad & W_{y_{1}}= & V_{0}^{p^{1}}\\
W_{x_{2}}= & V_{0}^{k^{1}} & \qquad & W_{u_{2}}= & V_{0}^{p^{2}}U_{p^{2}}^{0}V_{0}^{r^{1}} & \qquad & W_{y_{2}}= & V_{0}^{p^{2}}\\
W_{x_{3}}= & U_{0}^{q^{2}} & \qquad & W_{u_{3}}= & U_{0}^{\ominus p^{1}}V_{\ominus p^{1}}^{0}U_{0}^{k^{2}} & \qquad & W_{y_{3}}= & V_{0}^{p^{1}}\\
W_{x_{4}}= & U_{0}^{k^{2}} & \qquad & W_{u_{4}}= & U_{0}^{\ominus p^{2}}V_{\ominus p^{2}}^{0}U_{0}^{r^{2}} & \qquad & W_{y_{4}}= & V_{0}^{p^{2}}.
\end{array}
\end{equation}
Computing the terms of \eqref{eq:tdelay}, we obtain that $W_{u_{1}}W_{x_{2}}^{-1}W_{y_{2}}W_{y_{4}}^{-1}=V_{0}^{p^{1}}U_{p^{1}}^{0}=W_{y_{1}}W_{y_{3}}^{-1}W_{u_{3}}W_{x_{4}}^{-1}$,
an exact cancellation of the LHS, and the following results for the
parallel transported momenta:
\begin{eqnarray}
K_{1}^{b} & = & \hat{k}_{1}^{a}\left(V_{k^{1}}^{0}\right)_{a}^{b}=\sqrt{1+(4\pi G)^{2}k_{1}^{2}}\hat{k}_{1}^{b}=\cos\left(4\pi Gm_{1}\right)\hat{k}_{1}^{b}\nonumber \\
K_{2}^{b} & = & \hat{k}_{2}^{a}\left(U_{k^{2}}^{0}\right)_{a}^{b}=\sqrt{1+(4\pi G)^{2}k_{2}^{2}}\hat{k}_{2}^{b}=\cos\left(4\pi Gm_{2}\right)\hat{k}_{2}^{b}\nonumber \\
P_{1}^{b} & = & \hat{p}_{1}^{a}\left(U_{p^{1}}^{0}\right)_{a}^{b}=\hat{p}_{1}^{b}\nonumber \\
P_{2}^{b} & = & \hat{p}_{2}\left(V_{p^{2}}^{0}\right)_{a}^{b}=\hat{p}_{2}^{b}.
\end{eqnarray}
Each uppercase momentum only depends on the corresponding lowercase
one, and the torsion terms turn out to have no effect. The assumption
that $K_{1}$ and $K_{2}$ are parallel means simply that $\hat{k}_{1}=\hat{k}_{2}$.
The calculation of the time delay carries on exactly as done in \eqref{eq:tdelaynew}-\eqref{eq:tdelayfinal}
and the conclusion is immediate: since from \eqref{eq:masshellconst}
we know that photons are null, $P_{1}^{2}=P_{2}^{2}=\hat{p}_{1}^{2}=\hat{p}_{2}^{2}=0$,
and it results that there is no time delay between the intervals of
emission and reception in the 2+1 multiparticle model - as we expected
from \cite{key-2}, since the connection of momentum space is metric.
\\
\\
More surprising is the fact that the \textit{dual gravitational lensing}
effect derived in \cite{key-2} (angular deviation between the two
photons' momenta) is absent as well: since the decomposition \eqref{eq:photondecomp}
implies that $P_{1}\parallel P_{2}$ for null $P_{i}$ and $P_{i}=\hat{p}_{i}$,
we have $\hat{p}_{1}\parallel\hat{p}_{2}$. This is, however, compatible
with the result in \cite{key-2} for the deflection angle,
\begin{equation}
\Delta\theta=\frac{E_{1}+E_{2}}{2}\sqrt{\left|\eta_{ab}\left(T_{-}^{+}\right)^{a}\left(T_{-}^{+}\right)^{b}\right|},
\end{equation}
where $\left(T_{-}^{+}\right)^{a}=e_{-}^{c}e_{b}^{+}T_{c}^{ba}$,
$e_{a}^{+}$ being the null vector giving the direction of $P_{i}$
and $e_{-}^{a}=\eta^{ab}e_{b}^{+}$ the corresponding covector. Indeed,
it can be seen from \eqref{eq:2+1torsion} that
\begin{equation}
\left(T_{-}^{+}\right)^{a}\left(P_{i}\right)\propto e_{-}^{c}e_{b}^{+}\epsilon^{bad}\left(Ce_{c}^{+}e_{d}^{+}+\eta_{cd}\right)=0,
\end{equation}
so the predicted angular deviation vanishes in the 2+1 multiparticle
case.

\begin{onehalfspace}

\section{Conclusions}
\end{onehalfspace}

The main goal of the work described in this essay was three-fold:
to review the main physical concepts and mathematical formalism behind
the principle of relative locality, with particular focus on its application
to classical interactions between point particles (as a first step
to eventually study consequences of the new ideas in field theories
and ultimately quantum gravity); to examine an apparent problem of
the formalism, the ambiguities in defining the momentum constraints
of the classical action, and try to understand whether different definitions
produce different physics; and to give an example of how a well-studied
system with nontrivial phase space geometry and dynamics, the 2+1
multiparticle model in Einstein gravity, can be understood in terms
of the new framework.\\
\\
Towards the first goal, I tried to give a clear motivation and explanation,
based on intuitive ideas, of the material in \cite{key-1} and \cite{key-2}.
I believe that the main principle is not hard to grasp for anyone
with a basic understanding of differential geometry and special relativity,
as long as one gets used to thinking in momentum space - not an easy
task at all: in the words of Leonard Susskind, {}``only perverts
think in momentum space'', but in this case, it could be an useful
perversion.\\
\\
The second goal saw us making some considerations on the structure
of a vertex change, in particular identifying that identity-preserving
diffeomorphisms on a momentum constraint are the mark of a gauge transformation
involving the $\mathcal{K}^{(\alpha)}$ and interaction coordinates
$z^{(\alpha)}$, therefore just a redundancy in the physical description,
and doing several brute-force calculations with different conservation
laws based on the gamma delay experiment, which revealed one particular
instance where the structure of the derivation is altered - unless
torsion effects are not considered.\\
\\
Finally, regarding the third goal, we constructed a picture of phase
space that, despite being that of a classical gravity model, is remarkably
similar to the description of relative locality's phase space in the
limit $G_{N}\rightarrow0,\,\hbar\rightarrow0$ with a finite Planck
mass, to the extent where it could be studied using the same language
and methods. The 2+1 multiparticle model provided us with an interesting
example of a momentum space with a nontrivial but well understood
metric ($AdS_{3}$) and torsion, but no nonmetricity, and an interpretation
of the noncommutative nature of spacetime encoded in the relative
locality relations.\\
\\
As for possible further developments on this work, we did not, by
any means, exhaust the treatment of the change of vertex problem.
The first order results presented in Section 3.3 illustrated how the
analysis of the gamma delay problem seemed to change when the 2nd
and 3rd alternatives for momentum constraints were introduced, but
it was not completely clear whether that change produced an effective
modification in the physics - especially because the original calculation
presented in \cite{key-2} was done without torsion, and in this limit
all discrepancies disappear. A full calculation of the gamma delay
with torsion would probably have shed light on this issue. Alternatively,
going to second order and examining the curvature effects could prove
enlightening.

\section*{Acknowledgments}

I would like to express my gratitude, first and foremost, to my supervisor,
Laurent Freidel, for guiding me throughout the whole process of learning
and research, letting me work at my own pace and generally being an
excellent advisor.\\
\\
Thanks are also due to the whole PSI community, students, tutors and
administrative personnel alike, for being welcoming, supportive and
helpful at all times throughout the whole program, to all the people
with whom I had enlightening discussions throughout the making of
this essay, and to all that proofread it and helped me with the English
language and formal aspects.\\
\\
Last, but not least, a very special thank you goes to my parents,
for always being caring, supporting and motivating through my adventure
in Canada, and through all my life. I would not be here without you,
and I would not have been able to do this without you.

\section{Appendix - Review of 2+1 Einstein gravity}

For the purposes of this essay, we will introduce the \textit{first
order formalism} for Einstein gravity in 2+1 dimensions, which describes
it as a Yang-Mills-like theory with the Lorentz symmetry group $SO(2,1,\mathbb{R})\approx SL(2,\mathbb{R})$,
which Lie algebra is that of traceless $2\times2$ matrices, denoted
as $sl(2)$. A vector in this algebra can be written as $k=k^{i}\gamma_{i}$,
where $\gamma_{i}$ are the following matrices
\begin{equation}
\gamma_{0}=\left[\begin{matrix}0 & 1\\
-1 & 0
\end{matrix}\right],\,\,\,\,\,\,\gamma_{1}=\left[\begin{matrix}0 & 1\\
1 & 0
\end{matrix}\right],\,\,\,\,\,\,\gamma_{2}=\left[\begin{matrix}1 & 0\\
0 & -1
\end{matrix}\right].
\end{equation}
The dynamical variables describing geometry are then the frame fields
(\textit{dreibein}), which are derived from the natural projection
of the tangent bundle $e:\; TM\rightarrow M$,
\begin{eqnarray}
e_{a}^{i}(p):\: T_{p}M\times sl(2) & \rightarrow & \mathbb{R}\nonumber \\
(v,\, k) & \rightarrow & e_{a}^{i}v^{a}k_{i},
\end{eqnarray}
and the spin connection, which is the gauge group that induces sections
in $TM$,
\begin{eqnarray}
\omega_{a}^{i}(p):\: T_{p}M\times sl(2) & \rightarrow & \mathbb{R}\nonumber \\
(v,k) & \rightarrow & \omega_{a}^{i}v^{a}k_{i}.
\end{eqnarray}
From these we can construct one-forms in $T_{p}M$, $e^{i}=e_{a}^{i}dx^{a}$,
$\omega^{i}=\omega_{a}^{i}dx^{a}$ , as well as Lie-algebra valued
one-forms
\begin{equation}
\left[\begin{array}{c}
e\\
\omega
\end{array}\right]=\left[\begin{array}{c}
e_{a}^{i}\gamma_{i}dx^{a}\\
\omega_{a}^{i}\gamma_{i}dx^{a}
\end{array}\right]=\left[\begin{array}{c}
e_{a}dx^{a}\\
\omega_{a}dx^{a}
\end{array}\right]:T_{p}M\rightarrow sl(2).
\end{equation}
Spacetime metric is given by $g_{ab}=\eta_{ij}e_{a}^{i}e_{b}^{j}=\frac{1}{2}Tr\left(e_{a}e_{b}\right)$,
while the Levi-Civita connection is related to the spin connection
by the following formula, %
\footnote{Spacetime indices are raised/lowered with $g_{ab}$, while group indices
are raised/lowered with $\eta_{ij}$.%
}
\begin{equation}
\omega_{a}^{ij}=e_{b}^{i}e_{\,\,\,,a}^{bj}+e_{b}^{i}e^{cj}\Gamma_{ca}^{b},
\end{equation}
where the double-index notation for $\omega$ indicates the usage
of a different set of generators for $sl(2)$, $\omega_{a}=\omega_{a}^{ij}J_{ij}=\omega_{a}^{ij}\gamma_{[i}\gamma_{j]}$.\\
\\
The bulk Einstein-Hilbert-Palatini action for gravity can be rewritten
in terms of the new dynamical variables as
\begin{equation}
S=\frac{1}{16\pi G}\int_{M}tr\left(e\wedge F(\omega)\right),\label{eq:firstorderaction}
\end{equation}
where $F(\omega)=d\omega+\omega\wedge\omega$ is the curvature tensor
written in terms of the spin connection. The bulk equations of motion
are
\begin{eqnarray}
\frac{\delta S}{\delta e} & = & 0\Rightarrow F(\omega)=0\\
\frac{\delta S}{\delta\omega} & = & 0\Rightarrow de+[\omega,e]=0
\end{eqnarray}
and they state that geometry of spacetime in 2+1-dimensional vacuum
is flat. Hence, it becomes clear that three-dimensional gravity has
no local degrees of freedom - although it is possible to show that
boundary terms in the action lead to nontrivial topological dynamics.\\
\\
The action of Poincaré symmetry transformations on the dynamical variables
takes the usual form for a gauge theory:
\begin{eqnarray}
\text{Lorentz transformations:}\:\left\{ \quad\begin{array}{c}
\omega\rightarrow g^{-1}(d+\omega)g\\
e\rightarrow g^{-1}eg
\end{array}\right.,\, g & \in & SL(2),\\
\text{Translations:}\;\left\{ \begin{array}{c}
\omega\rightarrow\omega\\
e\rightarrow e+d\phi+[\omega,\phi]
\end{array}\right.,\,\phi & \in & sl(2).
\end{eqnarray}

\end{document}